# Molecular Spintronics

Masashi Shiraishi [1,2] and Tadaaki Ikoma [3,4]

1. Graduate School of Engineering Science, Osaka University, 1-3 Machikaneyama-cho, Toyonaka 560-8531, Japan.
2. PRESTO-JST, 4-1-8 Honcho, Kawaguchi 332-0012, Saitama, Japan.
3. Graduate School of Science and Technology, Niigata University, 8050 Igarashi 2-Nocho, Nishi-ku Niigata, 950-2181, Japan
4. CREST-JST, 4-1-8 Honcho, Kawaguchi 332-0012, Saitama, Japan.

E-mail) shiraishi@ee.es.osaka-u.ac.jp

**Abstract**

Molecular spintronics is recognized to as an attractive new research direction in a field of spintronics, following to metallic spintronics and inorganic semiconductor spintronics, and attracts many people in recent decades. The purpose of this manuscript is to describe the history of molecular spintronics by introducing important achievements and to show the current status of this field. In addition, the authors briefly introduce several theories for implementing studies in molecular spintronics.

# 1. Introduction

An electron possesses two degrees of freedom; charge and spin. Both degrees of freedom have been individually utilized in semiconductor physics and magnetism, yielding a tremendous amount of practical applications and expansion of horizons in basic science. Spintronics is though to be fusion domain of semiconductor physics and magnetism because both degrees of freedom in electrons are utilized simultaneously. In 1986, Gruenberg et al. reported the antiferromagnetic coupling of Fe in Fe/Cr/Fe layers, where magnetoresistance ratio was observed to be several % [1]. In 1988, Fert and co-workers reported the first giant magnetoresistance (GMR) of ~40% by using Fe/Cr multi-layers, which opened a door for this new research field, spintronics [2]. After their discovery, the similar GMR effects were observed in Co/Cr multi-layers [3] and so on. In 1992, a new system named granular was found where magnetic particles were dispersed in a non-magnetic matrix [4]. Co/Cu, Co/Ag and Co/Au are typical examples in the granular systems, and the magnetization direction of the ferromagnetic particles governs the spin transport phenomenon and the GMR ratio.

The GMR was a novel physical feature and the MR ratio was comparably high, so several practical applications, such as a magnetic reproducing head, were realized. In GMR, the Cr layers played roles of spin channel, namely, conducting spins were injected and transferred in the Cr. By contrast, insulating layers can be inserted between two ferromagnetic layers instead of conductive layers, which allows us to achieve spin-dependent tunneling transport. This is a tunneling magnetoresistance (TMR) effect. Miyazaki [5] and Moodera [6] individually observed the TMR effect in 1995, where the TMR ratio was up to 20% at room temperature (RT). They introduced an Al-O insulating barrier layer that was sandwiched by two ferromagnetic layers. Granular TMR was also reported [7] and unique temperature dependence of the tunnel resistance was revealed which was due to a charging effect [8]. In 2004, MgO single crystal was newly introduced in TMR devices as a tunneling barrier, which made coherent spin tunneling possible in a Fe/MgO/Fe system [9], and currently the TMR ratio is increased up to ~600% at RT [10].

Whereas the above history of spintronics is that of metallic spintronics, spintronics using inorganic semiconductors (GaAs, Si and so on) has been vigorously investigated. Among these inorganic semiconductors, GaAs could be used to realize spin transistors as proposed by Das and Datta [11], where injected spins into GaAs are rotated by an applied gate voltage because of existence of a strong spin-orbit interaction. On the contrary, Si is a comparably light element and has lattice inversion symmetry, and thus spin MOS field effect transistors (FETs) are expected to be realized. Currently, several research groups are investigating spin injection and spin transport intensively [12]. This field is known as the second pillar of spintronics, namely, semiconductor spintronics. Since 1999, a third pillar of spintronics attracts much attention from the people in spintronics and in molecular electronics, that is, molecular spintronics. Although detailed theoretical backgrounds will be described in detail in the next section, a molecule exhibits a comparably a smaller spin-orbit interaction. A spin-orbit interaction is known as the interaction which induced loss of spin coherence, thus, a material with a smaller spin-orbit interaction is needed to realize quantum computation systems and a so-called Sugahara-Tanaka type spin MOS FETs [13]. Currently, nano-carbonaceous molecules (graphene, carbon nanotube and fullerene) and organic molecules are intensively investigated for further progress in this research field, and a number important and attractive results has been obtained. Furthermore, spin-dependent tunneling transport via molecules also exhibits novel physical features that were not observed in metallic and inorganic semiconductor spintronics.

The purpose of this manuscript is to overview the history of molecular spintronics and to introduce important findings in the field. Section 1 is an introductory part, and Section 2 is mainly for introducing theories which should be understood in this field. Section 3 is for describing spin-dependent transport (spin injection and spin-dependent tunneling, namely, GMR and TMR effects) via molecules, and it consists of two parts; spintronics using nano-carbonaceous molecules and that using organic molecules. Section 4 describes a spin-dependent binding effect in organic molecules, which is observed with an external magnetic field but without ferromagnetic electrodes. Section 5 is the summary of the manuscript. One of the authors (T.I.) is responsible for Section 4, and the other author (M.S.) is responsible for the other sections.

2. Theories

In this chapter, several important concepts are introduced from a theoretical point of view. The first is a spin-orbit interaction. The spin-orbit interaction is a purely relativistic effect, which is derived from the Dirac equation in the natural unit,

$$i\frac{\partial}{\partial t}\psi(x,t) = [-i(\sum_{i=1}^{3}\alpha_i \frac{\partial}{\partial x_i}) + \beta m]\psi(x,t), \quad (1)$$

$$\alpha_i = \begin{pmatrix} 0 & \sigma_i \\ \sigma_i & 0 \end{pmatrix}, \sigma_1 = \begin{pmatrix} 0 & 1 \\ 1 & 0 \end{pmatrix}, \sigma_2 = \begin{pmatrix} 0 & -i \\ i & 0 \end{pmatrix}, \sigma_3 = \begin{pmatrix} 1 & 0 \\ 0 & -1 \end{pmatrix},$$

$$\beta = \begin{pmatrix} 1 & 0 \\ 0 & -1 \end{pmatrix},$$

where $t$ is time, $\psi(x, t)$ is a wave function, $x$ is a coordinate in a 3-dimensional space, $m$ is mass, $\alpha$ and $\beta$ are 4×4 matrices, $\sigma_i$ is a 2×2 Pauli's spin matrix and $i$ is an imaginary unit. Space coordinates and time are equivalent in relativistic quantum mechanics, and then the wave function possesses four components (3+1), where the upper two components of the wave function are for a positive energy solution, and by contrast, the lower two components are for a negative energy solution. In other words, the positive and the negative energy solutions are solutions for an electron and a positron, respectively. It should be noted that we can describe a spin degree of freedom by the Dirac equation as a spinor. Here, the wave functions for an electron and a positron are mixed in the

Dirac equation because of an existence of an off-diagonal $\sigma_t$ matrix in $\alpha_t$. In order to investigate interactions of the Dirac electrons in a non-relativistic limit, an appropriate digonalization of the Dirac Hamiltonian is needed. When we look back the eq. (1), the Dirac Hamiltonian is described as follows;

$$H = \begin{pmatrix} m \cdot I & \sigma \cdot p \\ \sigma \cdot p & -m \cdot I \end{pmatrix} = \alpha \cdot p + \beta m, \tag{2}$$

where $I$ is a $2 \times 2$ unit matrix and $p$ is momentum. Writing $U_F = e^{+iS}$ with $S$ hermitian and not explicitly time dependent, the unitary transformation is,

$$\psi' = e^{+iS}\psi, \tag{3}$$

$$i\frac{\partial}{\partial t}\psi = e^{+iS}H\psi = e^{+iS}He^{-iS}\psi' = H'\psi', \tag{4}$$

and $H'$ is to contain no off-diagonal components by construction. This transformation is called as the Foldy-Wouthuysen transformation. When we choose $U_F = e^{+iS}$ as,

$$e^{iS} = \exp(\beta\alpha \cdot p \cdot \theta(p)) = \cos|p|\theta + \frac{\beta\alpha \cdot p}{|p|}\sin|p|\theta, \tag{5}$$

the Dirac Hamiltonian under the Unitary transformation becomes,

$$H' = (\cos|p|\theta + \frac{\beta\alpha \cdot p}{|p|}\sin|p|\theta)(\alpha \cdot p + \beta m)(\cos|p|\theta - \frac{\beta\alpha \cdot p}{|p|}\sin|p|\theta). \tag{6}$$

Because $\alpha$ and $\beta$ are anti-commutable ($\{\alpha, \beta\}=0$),

$$\begin{aligned}
H' &= (\cos|p|\theta + \frac{\beta\alpha \cdot p}{|p|}\sin|p|\theta)(\alpha \cdot p + \beta m)(\cos|p|\theta - \frac{\beta\alpha \cdot p}{|p|}\sin|p|\theta) \\
&= (\alpha \cdot p + \beta m)(\cos|p|\theta - \frac{\beta\alpha \cdot p}{|p|}\sin|p|\theta)^2 \\
&= (\alpha \cdot p + \beta m)\exp(-2\beta\alpha \cdot p\theta) \\
&= \alpha \cdot p(\cos 2|p|\theta - \frac{m}{|p|}\sin 2|p|\theta) + \beta(m\cos 2|p|\theta + |p| \cdot \sin 2|p|\theta). \tag{7}
\end{aligned}$$

When we choose $\theta$ as,

$$\tan 2|p|\theta = \frac{|p|}{m}, \tag{8}$$

off-diagonal element, $\alpha$, can be eliminated, and

$$H' = \alpha \cdot p(1 - \frac{m}{|p|}\tan 2|p|\theta)\cos 2|p|\theta + \beta(m + |p| \cdot \tan 2|p|\theta)\cos 2|p|\theta$$

$$= \beta(m + \frac{|p|^2}{m})\frac{m}{\sqrt{p^2 + m^2}}$$

$$= \beta\sqrt{p^2 + m^2}. \qquad (9)$$

For further generalization, we introduce an electro-magnetic field,

$$H = \sigma(p - eA) + \beta m + e\phi \qquad (10)$$
$$= \beta m + \theta + \varepsilon,$$
$$\theta = \sigma(p - eA),$$
$$\varepsilon = e\phi,$$

where, $e$ is an electric charge, $A$ is vector potential and $\phi$ is scalar potential. Here, $\theta$ has off-diagonal elements. Remember that $\beta$ and $\theta$ are anti-commutable, whereas $\beta$ and $\varepsilon$ are commutable. We introduce the unitary transformation here again,

$$i\frac{\partial}{\partial t}e^{-iS}\psi = H\psi = He^{-iS}\psi' = e^{-iS}(i\frac{\partial}{\partial t}\psi) + (i\frac{\partial}{\partial t}e^{-iS})\psi',$$

and hence,

$$i\frac{\partial}{\partial t}\psi' = [e^{iS}(H - i\frac{\partial}{\partial t})e^{-iS}]\psi' = H'\psi'. \qquad (11)$$

The Foldy-Wouthuysen transformation is implemented three times with using the Baker-Hausdorff formula (see Appendix for the detail), and then the Hamiltonian can be written as,

$$H' = \beta\{m + \frac{(p - eA)^2}{2m} - \frac{p^4}{8m^3}\} + e\phi - e\frac{1}{2m}\beta(\sigma \cdot B)$$
$$- \frac{ie}{8m^2}\sigma \cdot \text{rot}(E) - \frac{e}{4m^2}\sigma \cdot (E \times p) - \frac{e}{8m^2}\text{div}(E). \qquad (12)$$

This is the Hamiltonian of the Dirac equation for an electron with an electro-magnetic field in a non-relativistic limit, and

$$-\frac{ie}{8m^2}\sigma \cdot \text{rot}(E) - \frac{e}{4m^2}\sigma \cdot (E \times p), \qquad (13)$$

is the spin-orbit Hamiltonian. When we assume spherically symmetric potential, $V(r)$, for simplicity, the first term of eq. (13) goes to zero. Finally, we obtain the spin-orbit Hamiltonian as,

$$H_{\text{spin-orbit}} = \frac{e^2}{4m} \frac{1}{r} \frac{\partial V}{\partial r} \sigma \cdot L, \qquad (14)$$

$$L = r \times p.$$

The above discussion is valid even when the atomic number, $Z$, is not equal to one (hydrogen), and it is notable that this Hamiltonian is proportional to $Z^4$ ($Z$: atomic number) under this assumption. Because the potential, $V(r)$, is a Coulomb interaction from a nucleus to a rotating electron around the nucleus, the spin-orbit Hamiltonian with a relativistic effect is described as,

$$H_{\text{spin-orbit}} = \frac{Ze^2}{4mr^3} \sigma \cdot L. \qquad (15)$$

When we calculate a classical spin-orbit Hamiltonian, the value is twice as large as the value in eq. (15), which is attributed to Thomas precession. Here, $r$ is inversely proportional to the atomic mass, $Z$, because a heavier atom possesses larger radii for outer-shell electrons, hence it can be understood that the spin-orbit interaction is proportional to $Z^4$ and light elements induce the smaller spin-orbit interaction [14].

Next, we discuss the conductance mismatch [15] and a spin drift/diffusion model is introduced for explaining the essence of the feature. An electric current, $J^{\text{charge}}$, can be written as follows;

$$\begin{aligned} J^{\text{charge}} &= J_\uparrow + J_\downarrow \\ &= \sum_{s=\uparrow\downarrow} (J_s^{\text{drift}} + J_s^{\text{diffusion}}) \\ &= \sum_{s=\uparrow\downarrow} (\sigma_s E + eD_s \text{grad}(n_s)) \\ &= \sum_{s=\uparrow\downarrow} (-\sigma_s \text{grad}(\phi) + eD_s \text{grad}(n_s)), \qquad (16) \end{aligned}$$

where $\uparrow, \downarrow$ indicates spin directions (up and down), $J^{\text{drift}}$ and $J^{\text{diffusion}}$ are a drift and diffusion current of spins, respectively, $n_s$ is a spin carrier density, $\sigma_s$ is conductivity ($=e^2 n_s \tau_s / m^*$, $\tau_s$ is momentum relaxation time and $m^*$ is an effective mass), $E$ is an electric field and $D_s$ is a diffusion constant. Here the Einstein's relationship is introduced and eq. (16) can be rewritten as,

$$J_s = \sigma_s \mathrm{grad}(\bar{\mu})/e, \qquad (17)$$
$$\bar{\mu} = \mu_s - e\phi,$$

where $\bar{\mu}$ is an electrochemical potential and $\mu$ is a chemical potential. In addition, differential equations for spin and charge are

$$\frac{\partial(\mathrm{spin})}{\partial t} + \mathrm{div} J_s = \frac{\sigma_s}{e}\frac{\bar{\mu}_\uparrow - \bar{\mu}_\downarrow}{l_{sf}^2}, \qquad (18a)$$
$$l_{sf} = (D_s \tau_{sf})^{1/2},$$
$$\frac{\partial(\mathrm{charge})}{\partial t} + \mathrm{div}(J_\uparrow + J_\downarrow) = 0, \qquad (18b)$$

where, $l_{sf}$ and $\tau_{sf}$ are a spin diffusion length and time, respectively. Eq. (18b) means charge current conservation, whereas eq. (18a) means that spins cannot be conserved because of spin relaxation. From above equations, spin diffusion equations become,

$$\Delta(\bar{\mu}_\uparrow - \bar{\mu}_\downarrow) = (\bar{\mu}_\uparrow - \bar{\mu}_\downarrow)/l_{sf}, \qquad (19a)$$
$$\Delta(\sigma_\uparrow \bar{\mu}_\uparrow + \sigma_\downarrow \bar{\mu}_\downarrow) = 0, \qquad (19b)$$

and eq. (19a) shows spin diffusion, and eq. (19b) shows an electric neutrality condition.

We apply the above equations for a practical case, namely, a heterojunction of non-magnet (NM) and ferromagnet (FM) without a tunneling barrier at $z=0$ (see Fig. 1). The electrochemical potential for up and down spins in the NM and the FM can be written as,

the NM side

$$\bar{\mu}_\uparrow = \frac{A}{\sigma_\uparrow}\exp(\frac{z}{l_{sf,F}}) + Bz + C, \qquad \bar{\mu}_\downarrow = -\frac{A}{\sigma_\downarrow}\exp(\frac{z}{l_{sf,F}}) + Bz + C, \qquad (20)$$

the FM side

$$\bar{\mu}_\uparrow = \frac{a}{\sigma}\exp(-\frac{z}{l_{sf,N}}) + bz, \qquad \bar{\mu}_\downarrow = -\frac{a}{\sigma}\exp(-\frac{z}{l_{sf,N}}) + bz, \qquad (21)$$

where $A, B, C, a$ and $b$ are constant, $l_{sf,N}$ and $l_{sf,F}$ are spin relaxation length in the NM and the FM, respectively. Then, the spin polarized current for each side can be calculated by using eq. (17), and the up-spin current in the FM side, for example, becomes,

$$eJ_\uparrow = \sigma_\uparrow \{ \frac{A}{\sigma_\uparrow l_{sf,F}} \exp(\frac{z}{l_{sf,F}}) + B \}. \qquad (22)$$

Because the electrochemical potential and the up- and down-spin currents are continuous at $z=0$, we can determine the constants ($A, B, C, a$ and $b$) with ease and obtain the final form of the spin polarized current at the interface between the NM and the FM as,

$$\frac{J_\uparrow - J_\downarrow}{J_\uparrow + J_\downarrow} = \frac{\beta}{1 + r_N/r_F}, \qquad (23)$$

$$\beta = \frac{\sigma_\uparrow - \sigma_\downarrow}{\sigma_\uparrow + \sigma_\downarrow} = \frac{\sigma_\uparrow - \sigma_\downarrow}{\sigma},$$

$$r_N = \frac{l_{sf,N}}{2\sigma},$$

$$r_F = \frac{l_{sf,F}}{4}(\sigma_\uparrow^{-1} + \sigma_\downarrow^{-1}),$$

where $\beta$ is the spin polarization of the conductivity, $r_N$ and $r_F$ is spin resistance in the NM and the FM, respectively. As one can see, the spin resistance is inversely proportional to conductivity, and if the spin resistance in NM (namely, the resistivity) is much larger than that in FM, eq. (23) goes to zero. This directly indicates that we can not expect spin injection into NM if the NM is highly resistive. Obviously, a metallic heterojunction is not the case. In the case of inorganic semiconductor and molecules, this conductivity mismatch becomes a serious problem for spin injection. It should be noted that the conductivity mismatch cannot be avoided even if one chooses an FM electrode with the same Fermi level as the levels of conduction or valence bands of a semiconductor. In such as case, we may avoid a Schottky barrier formation but we still face with the conductance mismatch because the conductivity of the FM and the NM is largely different.

There are several ways to avoid or solve this problem; one is insertion of a tunneling barrier (interfacial spin-dependent resistance) and the other is utilization of FM with lower conductivity, such as LaSrMnO (LSMO). Both ways are commonly used in spintronics, and here we show the effect of inserting a tunneling barrier from a theoretical point of view. Even when the tunneling barrier is inserted, the electrochemical potential for up- and down-spin is same as eqs. (20), and (21), and the spin polarized current can be written as same as in the previous case. The difference is manifested in the continuity condition of the electrochemical potential at $z=0$, namely, the electrochemical potential for up- and down-spin at $z=0$ is continuous when the spin-dependent interfacial resistance is included,

$$\overline{\mu}_\uparrow = \frac{A}{\sigma_\uparrow} + C + r_{i\uparrow} eJ_\uparrow = \frac{a}{\sigma}, \qquad (24)$$

$$\overline{\mu}_\downarrow = -\frac{A}{\sigma_\downarrow} + C + r_{i\downarrow} eJ_\downarrow = -\frac{a}{\sigma}, \qquad (25)$$

where $r_i$ is the spin-dependent interfacial resistance. The spin polarized current at FM/tunneling barrier/NM is described as,

$$\frac{J_\uparrow - J_\downarrow}{J_\uparrow + J_\downarrow} = \frac{r_F \beta + r_i \beta''}{r_F + r_N + r_i}, \qquad (26)$$

$$\beta'' = \frac{r_{i\uparrow}^{-1} - r_{i\downarrow}^{-1}}{r_{i\uparrow}^{-1} + r_{i\downarrow}^{-1}},$$

$$r_{i\uparrow} = 2(1-\beta'')r_i, \quad r_{i\downarrow} = 2(1+\beta'')r_i.$$

It should be noted that the spin polarization does not go to zero even when $r_N$ is much larger than $r_F$ if the interfacial spin-dependent resistance $r_i$ is comparable to $r_N$,

$$\frac{J_\uparrow - J_\downarrow}{J_\uparrow + J_\downarrow} = \frac{r_F \beta + r_i \beta''}{r_F + r_N + r_i} = \frac{(r_F \beta / r_i) + \beta''}{(r_F / r_i) + (r_N / r_i) + 1}$$

$$\approx \frac{\beta''}{(r_N / r_i) + 1} \text{ (if } r_N, r_i \gg r_F\text{)}. \qquad (27)$$

Eq. (27) clearly shows that the tunnel barrier plays a dominant role for realizing the injection the spin-polarized current into NM.

In the final part of this chapter, we briefly show the concept of a magnetoresistance effect by using TMR as an example. Figure 2 shows a schematic image of spin-dependent transport via a tunneling barrier. The resistance in parallel and anti-parallel magnetization alignments is written as

$R_P$ and $R_{AP}$, and then the MR ratio is defined as,

$$\frac{R_{AP} - R_P}{R_P} = \frac{2P_1 P_2}{1 - P_1 P_2}, \qquad (28)$$

$$P \equiv \frac{D_{Majority} - D_{Minority}}{D_{Majority} + D_{Minority}}.$$

where $P_{1,2}$ is spin polarization of each ferromagnet and $D$ is a density of state (DOS) of majority and minority spins. The magnetization alignments of the FM electrodes (parallel or anti-parallel) are controlled by applying an external magnetic field parallel to a long axis of the electrodes. The FM electrodes should have difference coercive forces by changing their geometries, and we can control the alignments in this manner. The spin polarization is determined by the difference of the DOS for up- and down-spin at the Fermi level. Because the polarization is often positive for ferromagnets, the MR ratio is often positive. However, when we change a bias voltage for a TMR device, the DOS at the pseudo-Fermi level can be also changed and the spin polarization can change its sign. This can induce the decrease/increase of the MR ratio or inverse of the sign of the MR ratio. Experimentally, the negative MR ratio was sometimes observed especially when LaSrMnO was used as one ferromagnetic electrode for molecular spin devices. This is thought to be ascribable to the inverse of the DOS at a finite bias voltage, and the bias voltage dependence of the DOS, namely, the sign of the MR ratio should be observed as the band structure changes. To certify this, a detailed analysis of the bias voltage dependence is demanded.

## 3. Spin-dependent transport via molecules

### 3.1 Various origins of magnetoresistance

Before introducing experimental results, we summarize various origins of MR except for spin-dependent transport, which are known to induce spurious signals. The most important is anisotropic MR (AMR), which is an effect that resistance of FM changes as a relative angle between an electric current direction and a magnetization direction changes. The AMR effect becomes obvious in 3d transition ferromagnetic metals and their alloys, for instance, NiFe, CoNi and FeCo [16], and the value of the MR ratio is typically 1-10%. AMR is the phenomenon that the electric current is modulated by the magnetization, and the electric current is induced by a motion of electrons. Because an electron has a spin degree of freedom, it is thought that the origin of the AMR is correlation between the spin and the motion of the charge, namely, the spin-orbit interaction. Experimentally, the AMR ratio exhibited clear dependence of a magnetic moment per atom [16], which is thought to be a direct proof. It is noteworthy that the AMR is found in every ferromagnet and we can not ignore this effect when we investigate spin-dependent transport. Before utilizing a new FM as an electrode, the measurement of the AMR should be carried out for a precise spin transport measurement.

As shown in Fig. 3, we assume that the magnetization and the electric current are not parallel. Then,

$$\begin{pmatrix} E_x \\ E_y \\ E_z \end{pmatrix} = \begin{pmatrix} \rho_\perp & -\rho_H & 0 \\ \rho_H & \rho_\perp & 0 \\ 0 & 0 & \rho_{//} \end{pmatrix} \begin{pmatrix} J_x \\ J_y \\ J_z \end{pmatrix}, \quad (29)$$

is a generalized relationship between the electric field and the current. Here, $\rho_\perp, \rho_{//}$ and $\rho_H$ are conductivity perpendicular to $z$-axis, that parallel to $z$-axis and the Hall conductivity, respectively. Alternating the relationship to the equation which describes resistance, $\rho = \mathbf{EJ}/J^2$,

$$\begin{aligned} \rho &= (E_x J_x + E_y J_y + E_z J_z)/J^2 \\ &= \rho_\perp + (\rho_{//} - \rho_\perp) J_z^2 / J^2 \\ &\equiv \rho_\perp + (\rho_{//} - \rho_\perp)\cos^2\theta, \end{aligned} \quad (30)$$

is obtained. Here, because no external magnetic field is applied and the magnetization of the FM is random, eq. (30) has to be averaged. Because $<\cos^2\theta>=1/3$, then,

$$\rho = \rho_\perp + (\rho_\parallel - \rho_\perp)/3$$
$$= \frac{2}{3}\rho_\perp + \frac{1}{3}\rho_\parallel \equiv \overline{\rho}. \qquad (31)$$

When $\mathbf{J} \parallel \mathbf{M}$, $\theta$ in eq. (30) is equal to 0, by contrast, when $\mathbf{J} \perp \mathbf{M}$, $\theta$ in eq. (30) is equal to $\pi/2$. From above discussion, the AMR ratio is described as,

$$\frac{\Delta\rho}{\overline{\rho}} = \frac{\rho_\parallel - \rho_\perp}{\overline{\rho}}. \qquad (32)$$

In a practical case for spin devices, each FM electrode can exhibit AMR. When resistance hysteresis is tried to be measured, the magnetization direction of both FM electrodes is set to be parallel in the beginning. This means that the magnetization of the FM is not random, but uniform. Next, an external magnetic field is applied in order to induce magnetization reversal in the FM electrodes. As described above, the resistance of the electrodes changes in the course of the application of the external magnetic field. When the magnetic field becomes large enough, the resistance comes back to the original value, because the uniform magnetization of the electrodes is recovered. In this process, resistance hysteresis can be observed in a so-called local measurement scheme [17], which is the reason why the contribution of the AMR should be eliminated by careful characterizations or by introducing the other experimental techniques. A non-local 4-terminal method is a potential method for excluding this spurious effect, and this is widely introduced in spintronics [17]. The detail will be described later.

The other origins of spurious signals are (1) local Hall effect, and (2) too large contact resistance between FM and molecules. The former occurs when the spin transport channel is not completely covered by a FM electrode (see Fig. 4) because the leakage magnetic field from the top-edge of the electrode affects the motion of spin carriers. The latter can induce large noise in spin signals and such a noise may be misunderstood as resistance hysteresis. Furthermore, reliability of the conventional "local" measurement scheme has not been fully clarified in molecular spintronics, which will be discussed in the following chapter, and the "non-local" 4-terminal measurement scheme is strongly needed when we certify the spin injection into new materials.

## 3.2 Molecular spintronics using nano-carbonaceous molecules

The door for molecular spintronics was opened by the first report on spin-dependent transport via multi-walled carbon nanotubes (MWNT) in 1999 [18]. Tsukagoshi and co-workers fabricated spin valve devices, where a MWNT acted as a spin channel. Figure 5 shows the data which they obtained, and the resistance hysteresis was observed up to 20 K. The splitting of the resistance hysteresis was not clear enough in some samples, because the difference of the coercive force in the Co electrode was not designed. Nonetheless, this is recognized as the first report on the spin transport via a molecule. Because a field of spintronics and that of carbon nanotubes started to garner much attention at that time, this report provides large impact as a pioneering study. After this study, many people started to try observing spin transport via MWNT and also via single-walled carbon nanotubes (SWNTs). However, not so much progress was not achieved, although several reports on MR via SWNTs and MWNTs [19-21]. One reason was difficulty in proving that spins were definitely injected into nanotubes, in other words, how to exclude possibilities of spurious signals such as AMR and how to clarify correspondence between resistance hysteresis and magnetization reversal of FM electrodes.

In order to solve this important issue, Sahoo et al. [22] and Man et al. [23] proposed a new approach. They reported gate-tunable spin transport via a SWNT and a MWNT by using PdNi ferromagnetic electrodes. There are two advantages in their study; (1) realization of low contact resistance by using the PdNi alloy, and (2) gate modulation of the channel conductance. The role of the PdNi electrode was to decrease contact resistance. Pd was famous for realizing an Ohmic contact to a SWNT [24] and the contact resistance between the FM electrode and the nanotubes was suppressed down to several k$\Omega$ when Pd was incorporated into Ni. Sahoo et al. firstly reported the low contact resistance in PdNi/MWNT [25], and this let us know the importance of the significance of realizing low contact resistance. The gate modulation of the conductance induced Coulomb blockade in the nanotube, and then a clear Coulomb oscillation was observed, as was observed previously [26]. Hence, on- and off-resonant states were generated in the SWNTs, where the conductance of the nanotubes was oscillatory controlled. Based on the Landauer formula in a ballistic regime, the correspondence between the oscillation of the conductance and the MR ratio observed in the nanotube was theoretically explained, which was recognized as a proof of spin injection and transport in the nanotube (Fig. 6).

In 2007, an importance milestone in molecular spintronics was realized, which is generation of a pure spin current in a molecule [27]. A pure spin current is completely different from a charge current in its concept. A pure spin current is a current of spin angular momentum, and there is no charge current there (see Fig. 7).As shown in Fig. 7, the same number of up- and down-spin flows to the opposite side each other, and there is no charge flow. However, the down spin current to the left side is equivalent to the up spin flow to the right side. Hence, a pure spin current ideally possesses time-reversal symmetry and is a dissipationless current (experimentally, a diffusive pure spin current is often generated, which is a dissipation current). A pure spin current was already realized in metallic spintronics by using Co/AlO/Al spin valves [28]. Jedema et al. reported MR in the Co/AlO/Al spin valve by introducing a non-local 4-terminal method (Fig. 8). Because spurious effects are eliminated in this method, the non-local method allows us to obtain reliable results and to implement valuable discussions. Currently, the non-local method is widely used to prove spin injection into condensed matters, such as non-magnetic metals, Si, GaAs and molecules, and plays an indispensable role to obtain reliable results [29]. The study by Tombros provided clear proof of the spin injection into a SWNT at 4 K (Fig. 9), and also called people's attention to the reliability of the conventional local 2-terminal method. The spin signal intensity in the non-local method should be a half of that in the local method because difference of electrochemical potential that is experimentally observed in the non-local method is only for up- or down-spin while the difference in the local is for both (Fig. 10) [30]. Whereas this was experimentally observed in metallic spin valves [30], the intensity in the local method observed in their study was 35 times larger than that in the non-local, and the authors pointed out the importance of clarification of the reliability of the local method. In other words, their study taught us that we cannot conclude spin accumulation and spin injection as far as we use the local method only. In this sense, it should be emphasized that (1) MR-induced resistance hysteresis, (2) generation of non-local pure spin current and (3) Hanle-type spin precession [31] (described later) should be observed when we try to prove spin injection.

The other important milestone in nanotube spintronics is an observation of very large MR ratio by using a LSMO/MWNT/LSMO spin valve [32]. The authors introduced two LSMO electrodes, which were expected as half-metallic electrodes at low temperature, and observed the MR ratio of ~60% at 5 K and estimated the spin relaxation length to be 50 μm assuming the MWNT mean free path of ~100 nm.

Based on the progress of experimental techniques and accumulation of understandings how to obtain a reliable result in nanotube spintronics, graphene spintronics was brilliantly established and has been dramatically developed since 2007. Graphene is currently one of the hottest materials in solid state physics, because a number of interesting and unique physics is realized in it. Integer quantum Hall effect at room temperature [33], fractional Hall effect at low temperature [34], a proximity effect [35] and field effect transistors with an extraordinarily large mobility [36] have been realized, because the zero-mass Dirac fermion system can be realized in graphene. Introduction of a spin function in graphene was another attractive challenge, and reliable and reproducible results on spin injection and generation of a spin current in single- and multi-layer graphene (SLG and MLG) at room temperature were firstly reported by several groups individually [37-39], which was achieved by introducing the non-local method. The advantage of the non-local method was shown in the literature [37], because AMR effect observed in the local method was fully eliminated in the non-local method (Fig. 11). Furthermore, the Hanle-type spin precession was observed in SLG [38] and MLG [40], which is the direct proof for the spin injection (Figs. 12(a) and (b)). In order to generate the precession, we apply an external magnetic field perpendicular to the spin transport channel (in this case, graphene) (see Fig. 13). The in-plane spins, which are injected into graphene, starts to process in the course of standing up parallel to the magnetic field. The spins diffuse to the detector electrode (Co2) with precessing. The output signal depends on the relative angle between the transferred spins and the detector spins, namely, the signal oscillates as the spins precess. The oscillation and a crossing of the spin signals in parallel and anti-parallel magnetization alignments allow us to conclude the precession of the injected spins. The non-local resistance with the precession can be described as,

$$\frac{V_{non-local}}{I_{inject}} = \frac{P^2}{\sigma A/D} \int_0^\infty \frac{1}{\sqrt{4\pi Dt}} \exp(-\frac{L^2}{4Dt}) \cos(\omega t) \exp(-\frac{t}{\tau_{sf}}) dt, \qquad (33)$$

where $P$ is spin polarization, $\sigma$ and $A$ are conductivity and a crosssection of the channel, $D$ is a diffusion constant, $L$ is a gap length between two FM electrodes, $\omega$ is the Lamour frequency (= $g\mu_B B/\hbar$), $t$ is time and $\tau_{sf}$ is the spin coherent time. The analytical solution of eq. (33) was described in the literature [41],

$$\frac{V_{non-local}}{I_{inject}} = \frac{P^2 \lambda_N}{2\sigma A} \exp(-\frac{L}{\lambda_N})(1+\omega^2 \tau_{sf}^2)^{-1/4} \exp(-\frac{L}{\lambda_N})\{\sqrt{\frac{1}{2}(\sqrt{1+\omega^2 \tau_{sf}^2}+1}-1\}$$
$$\times \cos\{\frac{\tan^{-1}(\omega\tau)}{2}+\frac{L}{\lambda_N}\sqrt{\frac{1}{2}(\sqrt{1+\omega^2 \tau_{sf}^2}-1)}\}. \qquad (34)$$

By using the equation, a couple of important spin transport parameters (diffusion constant, spin coherent time and spin polarization) are estimated, and spin coherent length, $l_{sf}$ is also calculated from the following relationship,

$$l_{sf} = \sqrt{D\tau_{sf}}. \qquad (35)$$

An example is shown in Figs. 12(a) and (b). It is notable that one can also estimate spin relaxation length by using results on length dependence of spin signals, but the estimation is verified only in the non-local regime because it was pointed out that all of spin signals in the local regime is not due to spin accumulation as mentioned above. As far as investigated, the spin coherent length and time in graphene is typically 1-2 μm and 100-200 ps, which is currently apart from the values in a theoretical prediction and should be improved in the future.

A number of studies has been implemented in graphene spintronics, so far, and anisotropic spin relaxation [42], spin drift in graphene [43, 44], correspondence between carrier densities and spin coherent time [45], unprecedented robustness of spin polarization in graphene [40], electron-hole asymmetry in spin transport [46], modification of spin transport by a chemical doping [47], investigation of tunneling barrier [48, 49], demonstration of gate-induced modulation of spin signals [39, 50-52] and so on. Here, the anisotropic spin relaxation is introduced as one example of an attractive spin transport character in graphene. They injected spins perpendicular to the graphene plain by applying the external perpendicular magnetic field of above 1.5 T, and compared spin relaxation time under B=0 T and 2 T. Then, they found that spin relaxation of the spins perpendicular to the graphene plain is 20% faster than that parallel to the plain, which was attributable to the fact that the dominating spin relaxation mechanism is the Elliot-Yafet mechanism (Fig. 14). Concerning the correspondence between spin signal intensities in the local and non-local schemes, which has been unclear in molecular spintronics as described above, it was clearly verified and steadfast basis of graphene spintronics was constructed in graphene spintronics (Fig. 15) [40], and thus graphene is recognized as a model material for discussing spin transport in molecular materials. It is also notable that a new approach for spin injection into graphitic materials was recently reported by Banerjee et al. [53], where perpendicular spin transport was realized by employing ballistic electron magnetic microscopy (Fig. 16).

Contrary to studies on spintronics using graphene and carbon nanotubes, spintronics using fullerene has not been intensively studied until recently. The first report on spin-transport via $C_{60}$ was implemented by fabricating a granular structure [54], in which Co particles (4-5 nm in diameter) were dispersed in various matrix species. Micklitz and co-worker observed MR up to 60 K, and the MR ratio was ca. 30% at 4 K. They concluded that the spin transport was governed by TMR. The interesting feature in their study was the TMR was observed in Co/CO or Co/Xe systems. However, it was not clarified that the MR was attributed to the magnetization of the Co and whether the MR was able to be observed up to higher temperature. Based on this motivation, Miwa et al. demonstrated the MR up to room temperature by using a $C_{60}$-Co nano-composite system that has a similar structure with that of granular, and they also clarified that the observed MR was definitely due to the magnetization alignments of the Co nano-particles of 2-3 nm in diameter [55]. The correspondence of the MR and the magnetization was clarified in the same sample, where the MR curve and the square of the normalized magnetization coordinated very well (Fig. 17). Sakai et al. also reported the similar effect by using the $C_{60}$-Co system but the observed MR ratio was extremely large (80% at 2 K in their definition, but ~300-500% in the conventional definition, see Fig. 18) [56, 57]. They investigated the electronic structure of the $C_{60}$-Co covalent state by controlling the contents of the molecule and found that the spin polarization of the Co was strongly enhanced at the interface between $C_{60}$-Co by using X-ray magnetic circular dichroism [58]. It is noteworthy that recently Shiraishi and co-workers reported gigantic TMR ratio of hundred-thousand % in the $C_{60}$-Co nano-composite at 2 K [59]. It was not so much expected that a new physical feature would be found in spin-dependent tunneling transport via molecules at the first stage of such studies, but unexpected interesting issues have been observed, and nowadays, it is recognized that it is forth for implementing further investigation is needed. The other important findings in TMR via organic molecules will be described in the next subsection.

### 3.3 Molecular spintronics using organic molecules

In this subsection, spin-dependent transport via organic molecules is introduced in detail. In 2002, Dediu and co-workers reported the change of the resistance of sexithienyl (T6) under an application of an external magnetic field by using LSMO electrodes (Fig. 19) [60], which is recognized as the first report on an MR effect using an organic molecule. Following to this work, optical detection of spin coherent transport via molecularly bridged quantum dots was reported by Ouyang et al. in 2003 [61]. Because electrical detection of spin transport is comparably easier than optical detection in this early stage and Dediu used a lateral structure without the difference of the coercive force of the LSMO, Xiong et al. improved a device structure by introducing a sandwiched structure (Co/Alq3(~130 nm)/LSMO) and observed an MR effect with resistance hysteresis in 2004 (Fig. 20) [62]. The MR effect was observed up to ~250 K which was close to the Curie temperature of LSMO. After these pioneering studies, a theoretical study in which large MR ratio of several hundreds % was predicted was carried out [63] and organic spintronics attracted much attention [64-66].

Contrary to nano-carbon spintronics, the non-local spin injection, generation of a pure spin current and the Hanle-type spin precession have not been achieved in organic spintronics. However, the unique investigations from the other stand-points were implemented [67, 68]. One is spin-resolved two-photon photoemission [67]. Cinchetti et al. fabricated a CuPc/Co heterojunction, and injected spin polarized photons from the Co into the lowest unoccupied level of the CuPc (the spin injection efficiency of 85-90%). They reported the spin flip length of several tens nano-meters in the CuPc at room temperature (Fig. 21). Although this is not electrical detection of spin injection, it is valuable that this clarified the precise spin flip length in amorphous molecules. The other important study was carried out by using muon spin resonance [68], in which Drew and co-workers fabricated a NiFe/LiF/Alq3/TPD/FeCo vertical spin valve device. They observed MR in this device and simultaneously carried out depth resolved investigations for detecting spin polarization, which allows us to obtain a reliable proof of spin injection into organic molecules. According to their study, the spin diffusion length in the Alq3 was typically several tens nano-meters at low temperature (Fig. 22). These two studies pointed out that the spins which were injected both optically and electrically made relaxation less than 50 nm, and the results do not have accordance with previous studies using electrical techniques for spin injection and thicker (longer) spin transport channels. Here, it is noteworthy that spin transport even in a short organic spin channel using rubrene, Alq3 and CuPc (<10 nm) is often tunneling spin transport (for example, see Fig. 23) [69-73] and tunneling spin transport can be realized via ferromagnetic particles which were unintentionally formed in organic spacers [74], and also that Jiang et al. reported the absence of the spin transport in the case of hole injection in Alq3 sandwiched by Co and LSMO [75]. On the other hand, Fahlman and Dediu carried out careful studies on an investigation of a band alignment of Co/Alq3/LSMO by introducing photoemission spectroscopy (Fig. 24) [76-78] and clarified that spin transport in the lowest unoccupied molecular orbital can induce spin-dependent MR, because Alq3 is known as an appropriate electron transport material in a field of organic electroluminescence devices. Furthermore, they realized an optimum quality of an interface between Alq3 and $Al_2O_3$ which is a key issue to avoid the conductance mismatch for spin injection into organics and achieved spin transport via Alq3 at room temperature (Fig. 25) [79]. Here, it should be emphasized that a hyperfine interaction plays an important role for spin relaxation in organics, and Bobbert theoretically clarified the role of the hyperfine interaction in Alq3 for the spin transport and the shape of the resistance hysteresis [80]. The author showed the good accordance of the theoretical fitting and the experimental data obtain by Dediu, which can be a proof of spin injection into Alq3 in their spin valves. Although negative MR was observed in their Alq3 spin valves, which does not seem usual based on a conventional spin transport model, they concluded in the article that the negative MR was ascribable to the combination of DOSs in Co and LSMO, namely, the DOS of the minority spins in the FMs was dominant in the spin transport (see Fig. 25). It is noteworthy that a multi-functional spin device was fabricated by using this Alq3-based spin valve [81].

As described above, it is not still conclusive whether GMR or TMR governs spin transport via organic molecules because there is no report on the Hanle-type spin precession. Of course, occurrence of GMR in organics is strongly expected, but it has been understood that TMR via organics includes interesting physical aspects. One example is an observation of large TMR ratio of ~300% at 2 K via Alq3 (Fig. 26), where a vertical spin device was fabricated by a nano-indentation method [82]. Positive MR was found in their study, contrary to the previous work [79], and they clarified that the formation of a hybridized state between the Alq3 and the FM induced the positive MR. Currently, the MR ratio is much more enhanced when they utilize CuPc as a spacer [83]. By contrast, Sun et al., reported large negative MR of -300% at 10 K in Co/Alq3/LSMO vertical spin valves [84]. Concerning large MR ratio, Sakai et al. recently reported MR ratio of ~100,000% at 2 K by using a rubrene-Co nano-composite structure [59], and further investigations for clarifying the underlying physics of such the large MR ratio is expected. The other interesting feature in TMR via molecules was reported by Hatanaka et al., where Co nano-particles were uniformly embedded in a rubrene matrix and higher order (at the maximum $5^{th}$-order) co-tunneling spin transport was realized [85]. Second order co-tunneling was already observed in insulating granular systems and theoretical explanation was clearly done [86], where MR ratio was enhanced due to Coulomb blockade and appearance of the co-tunneling effect. Hatanaka et al. observed the other type of the enhancement of the MR ratio inside of the Coulomb gap (see Fig. 27), which was not able to be explained by the model by Takahashi and Maekawa [86]. They investigated *I-V* curves inside of the Coulomb gap and found that the power of the bias voltage was at maximum 9, which indicated that the $5^{th}$ order co-tunneling occurred and this induced the further enhancement of the MR ratio up to 80%. It was also clarified by using a $^{57}$Co spin echo method that the origin of such the large MR ratio observed in molecular nano-composites was enhancement of spin polarization between rubrene and Co [87]. Whereas these studies were implemented by using the nano-composite structure, in which ferromagnetic nano-particles were dispersed, the similar structure was realized by using $Fe_3O_4$ nano-particles. Wang et al. reported an observation of an MR effect in a combination of $Fe_3O_4$ and oleic acid molecules [88].

In the final part of this subsection, several interesting topics are introduced. The first one is detection of circularly polarized light from an organic light emitting diode with FM electrodes [89]. This is probably only one report on the detection of the spin-polarized luminescence in our best knowledge. The fusion of a spin degree of freedom and photon could give many opportunities to fabricate novel spin devices using molecules, and so further investigation is awaited. The second one is utilization of molecular magnet, V(TCNE) for molecular spin devices with a short spin channel (~10 nm) [90]. The Curie temperature of V(TCNE) was reported to be ~400 K, which overcome the limitation of the Curie temperature of LSMO that is often used for molecular spin devices. In the conductance-mismatch-based point of view, FMs with low conductivity is a hopeful material for avoiding the conductance mismatch. Hence, the introduction of molecular magnets provides us a new point of view for the discussion of spin injection and transport in organics. The third one is an isotope effect [91]. Nguyen et al. fabricated vertical spin valves using a polymer material (DOO-PPV) with and without replacing hydrogen (H) to deuterium (D) in the polymer. The structure of the spin device was asymmetric, Co/DOO-PPV(25 nm)/LSMO. According to their experiments, MR ratio was increased up to ~40% at 10 K in the device with deuterium because the hyperfine interaction was thought to be suppressed, which is the isotope effect, and the bias voltage dependence of the MR ratio was symmetric.

## 4. Spin-dependent binding in molecules

Not only the transport but also the binding process of the carriers depends on the spins in molecules, which results in organic magnetoresistance (OMAR) effect. It is noteworthy that the OMAR effect is an intrinsic property of organic semiconductor from the fact that it is observed in organic devices without any magnetic electrodes and materials. This section is devoted to introduce some interesting phenomena in molecular semiconductors due to spin-dependent binding process, which have different origins from the transport-related topics described closely in the previous sections.

In organic semiconductors there possibly exist many kinds of quasi particles, such as electron, hole, polaron, exciton, soliton and so on. Since singly charged particle that is usually a promising carrier in organic semiconductor has an unpaired electron spin, the spin state of the carrier is basically doublet (D). The D state is the case for electron, hole and polaron. The excitons may also have unpaired electrons although it is normally neutral. The spin state of the Frenckel type of exciton that is a major exciton in organic semiconductor is singlet (S) or triplet (T). These quasi particles can encounter with other particles during transport or diffusion, and pair off with them. The particle pair forms new spin states. Fig. 28 shows the spin states for the possible combination of D and T particles. Because pairing with the S-particle, which can also take place among the mobile particles, does not change the total spin multiplicity at all, it is not directly connected with the magnetic field effects on spin dynamics. In this section we shall restrict ourselves the DD pair among three cases shown in Fig. 28, because the DD pair mechanism seems to be deeply associated with the OMAR observed recently. The core of the DD pair mechanism is the competitive processes among the spin conversion dynamics ($k_{ij}$, $i \neq j$) in the pair and the spin conservative decays ($k_i$) from the pair.

## 4.1 Doublet-doublet pair mechanism

Most of electrons, holes and polarons that are main charged carriers in organic semiconductor have a D state. The pair with those two particles with the D state forms the S and T states, depending on the mutual orientation between the electron spin angular momenta (Fig. 29(a)). We call this type of pair as the DD pair ([DD]) as shown in Fig. 28(a). The DD pair may have a binding process to $X$:

$$D + D \quad \overset{1,3}{\rightleftharpoons} [DD] \xrightarrow{p(spin)} X, \qquad (36)$$

such as recombination or bipolaron formation. If the binding probability ($p$(spin)) depends on the spin multiplicity of the DD pair or the rate constants ($k_S$ and $k_T$) from the S and T states are not equivalent, the MR effect can be induced. This is because the population ratio among the four spin sublevels of the DD pair can be modulated by the Zeeman interaction with an external magnetic field ($B_0$). The population balance is determined by the spin conversion rate constant between the S and T states ($k_{ST}$) which is a function of the $B_0$ field. Therefore the mechanism for the spin conversion is a key issue to understand the MR effect due to the DD pair.

The spin Hamiltonian $\hat{H}_0$ of the DD pair, which is deeply connected with the spin conversion, comprises the exchange ($\hat{H}_{ex}$), Zeeman ($\hat{H}_Z$) and hyperfine ($\hat{H}_{hf}$) interactions.

$$\hat{H}_0 = \hat{H}_{ex} + \hat{H}_Z + \hat{H}_{hf}. \qquad (37)$$

The $\hat{H}_{ex}$ interaction with a coupling constant of $J$ causes electron spin exchange at a frequency of $2J/\hbar$ and splits the S and T states as illustrated in Fig. 30. If the exchange frequency or the energy splitting of $2|J|$ exceeds the other interactions of $\hat{H}_Z$ and $\hat{H}_{hf}$, each electron spin will experience the same local magnetic field ($B_L$), so that any difference of Larmor frequency of electron spins does not exist and no spin conversion occurs. Because the $\hat{H}_{ex}$ interaction is effective only in a range where the electron orbitals overlap, it is often expressed with a rapidly decreasing function of the separation distance between the two D particles ($r$): $J_0\exp(-\alpha r)$. Hence it is a time-dependent interaction by the particle motion. The $J$ value becomes negligible at the long $r$ distances than the critical distance ($r_c$) that is usually shorter than the Onsager distances of a few nanometers for electron-hole pair in organic semiconductor. When the $B_0$ field is not applied, all the spin levels are degenerate as depicted in Fig. 28(a) because of $\hat{H}_0$ = $g\mu_B B_0$ = 0. Therefore, the hyperfine interactions are dominant in $\hat{H}_0$ of the DD pair in the regime-A of Fig. 30(a) and play an essential role for the effective spin conversion in the DD pair.

The hyperfine interaction is a magnetic interaction of electron spin ($S$) with nuclear spin ($I$):

$$\hat{H}_{hf} = \sum_i^{N_1} \hat{S}_1 \mathbf{A}_i \hat{I}_{1i} + \sum_j^{N_2} \hat{S}_2 \mathbf{A}_j \hat{I}_{2j}. \qquad (38)$$

$N$ is the number of magnetic nuclei interacting with each D particle, such as $^1$H ($I$ = 1/2), $^{14}$N ($I$ = 1), $^{35,37}$Cl ($I$ = 1/2), $^{31}$P ($I$ = 1/2) and $^{19}$F ($I$ = 1/2) that have high natural abundance more than 99 % and are important elements for functionalizing molecular semiconductors by bonding to carbon or replacing in molecular framework. The coupling matrix of $\mathbf{A}$ involves Fermi contact and magnetic dipolar interactions [92,93]. There are a lot of non-negligible matrix elements of $\hat{H}_{hf}$ among the spin sublevels of S, T$_-$, T$_0$ and T$_+$ [94,95]. The nonzero off-diagonal elements of $\langle i|\hat{H}_{hf}|f\rangle$ in $i \neq f$ are a driving force for the coherent spin conversion between the S and T states. This type of spin conversion is identical to the mutual orientation alternation of the electron spins due to the periodical precession of individual electron spins in the vector model of Fig. 29(b). The precessional axes for two D particles are the local magnetic fields of $B_{1L}$ and $B_{2L}$ that are mainly caused by the nuclear spins of the particle 1 and 2 in the DD pair respectively. In general, the $B_{1L}$ and $B_{2L}$ have different direction and magnitude at $B_0 = 0$ because the same nuclear spins does not necessarily interact with the two component D particles. The $B_0$ field acts as an additional common field for the particles in the DD pair and also tends to reorient the effective fields of $B_{1,2}$ (= $B_{1L,2L} + B_0$) to the same direction as shown in Fig. 29(c). Hence the electron spins precess around the similar axes, which may result in the increase of the precessional amplitude and angular velocity. The stronger correlation between the D particles due to the common precessional axis, in other word common quantization axis, enhances the spin conversion and eventually increases the $k_{ST}$ constant in ultralow field region as shown in Fig. 30(b).

More drastic magnetic field effect on the spin conversion occurs in the field region more than the hyperfine coupling constant. Not only the $B_0$ field makes the spins of the D particles precess around the same axis (Fig. 29(a)) but also the Zeeman interaction shifts the $T_+$ and $T_-$ sublevels in energy (Fig. 30(a)). The $\hat{H}_Z$ term has the diagonal element of $|\langle T_\pm | \hat{H}_Z | T_\pm \rangle|$. The energy shift of the $T_+$ and $T_-$ sublevels prohibits the transitions between the S and $T_\pm$ levels because of the relative reduction of the off-diagonal elements of $|\langle T_\pm | \hat{H}_{hf} | S \rangle|$ versus the diagonal elements of $|\langle T_\pm | \hat{H}_Z | T_\pm \rangle|$. In consequently, the $k_{ST}$ gradually decreases with the increase of $B_0$ and saturates in the high fields as shown in Fig. 30(b). The saturated $k_{ST}$ value at the high fields of $\hat{H}_Z$ $\hat{H}_{hf}$ where the outer triplet sublevels are cut off from the spin conversion arises from the transition between the S and $T_0$ states that maintain the same energy as at zero field. The saturation behavior may be characterized by a field value of $B_{1/2}$ defined by the next equation [96], which is a field where half of the saturation effect is obtained.

$$B_{1/2} = \frac{2\left(B_{1hf}^2 + B_{2hf}^2\right)}{B_{1hf} + B_{2hf}}, \qquad (39)$$

$$\text{where } B_{ihf} = \sqrt{\sum_j I_{ij}\left(I_{ij} + 1\right)a_{ij}^2}. \qquad (40)$$

There is another interesting region where the S state can energetically meet with one of the three T sublevels, which is the crossline-R in Fig 30(a). The degeneracy of the S and T. states comes from the non-vanishing $\hat{H}_{ex}$ term in the DD pair with short $r$ distance, which induces an energy separation between the S and T sublevels even at zero field. Because of the S-T energy gap, the spin conversion does not take place effectively at zero field. However the Zeeman interaction is able to compensate the energy gap at certain fields where a condition of $2|J| = g\mu_B B_0$ is fulfilled. Therefore the spin conversion rate of the short distance DD pairs is expected to give a resonant type of magnetic field dependence as depicted Fig. 30(c). The rate constant increases near a $B_0 = 2|J|/(g\mu_B)$, where the occasional degeneracy allows the spin conversion between the S and T. states due to the hyperfine interaction. The resonance between the S and T$_+$ states is also possible if the DD pair has a positive $J$ value. However it should be noted that MR due to the resonant type of dependence may be hardly detected because staying time of the DD pair at the narrow region of the crossline-R is very short normally.

The spin conversion mentioned above is based on the coherent spin dynamics. However incoherent spin dynamics that is spin relaxation also play an important role for the spin conversion. Within the framework of the Redfield theory [97,98], the spin relaxation of individual D particle is caused by a fluctuating local field ($\delta B$) characterized with a correlation time ($\tau_c$).

$$\hat{R}_1 = -\left(\frac{g\mu_B}{\hbar}\right)^2 \overline{\delta B_\perp^2} \frac{\tau_c}{1+\omega_0^2 \tau_c^2}, \qquad (41)$$

$$\hat{R}_2 = -\left(\frac{g\mu_B}{\hbar}\right)^2 \left\{\overline{\delta B^2}\tau_c + \frac{1}{2}\overline{\delta B_\perp^2}\frac{\tau_c}{1+\omega_0^2 \tau_c^2}\right\}, \qquad (42)$$

$$where\, \omega_0 = \frac{g\mu_B B_0}{\hbar}. \qquad (43)$$

The relaxation operators of $\hat{R}$ clearly are a function of $B_0$. Anisotropy of the $g$ and hyperfine tensors, which are due to the spin-orbit and electron spin-nuclear spin dipolar interactions respectively, are important origin for $\delta B$. These interactions can be time-dependent local magnetic fields by random orientational change of the D particle due to thermal molecular librations and/or lattice vibrations. At low magnetic fields where the organic MR effect appears, we cannot expect the relaxation effect due to the anisotropic $g$ factor because of intrinsic small spin-orbit coupling of organic semiconductor, but anisotropy of hyperfine can produce effective relaxations. In addition, as far as the mobile D particle in solid state semiconductors concerned, the isotropic hyperfine interaction also can induce the spin relaxation since the value of coupling constant varies with time as the result of the charge transfer from one molecular unit to others. For the DD pair of which the $r$ distance changes with time, furthermore, there are other mechanisms for the spin relaxation. One comes from the electron spin-electron spin dipolar interaction between the D particles that is omitted in eq. 37 for simplicity, the other from the exchange interaction. Because the dipolar and exchange interactions are $r$-dependent, the relaxation effects may be significant in the short distance DD pairs rather than in long distance pair.

## 4.2 MR based on the DD pair mechanism

One of the earliest studies related with the MR effect in organic semiconductor, which is based on the DD pair mechanism, may be a series of researches of magnetic field effect on photoconductivity by Frankevich and coworkers using crystal and films of anthracene and tetracene [99-101]. Frankevich has explained the magnetoconductance of photocurrent in terms of different *g* factors of hole and electron, operating on a loosely bound charge DD pair that might be Wannier exciton [102]. Magnetic field effects due to the DD pair mechanism seem to be more clearly observed in interface dynamics between different organic semiconductors [103-106], because the electron-hope pair, which might be sometimes called charge transfer exciton, is dominant species at the interface just after photoexcitation. For example, the intensity of the exciplex photoluminescence of electron-hole pair in the thin tetracene films on anthracene is reduced by a few percent by applying magnetic field of a few mT [103]. The $B_{1/2}$ value of this magnetic field dependence of the photoluminescence is 1.6 mT. The electron-hole pair created by the photoinduced primary electron transfer from one organic crystal to the other solid at the interface is a precursor to not only generate carriers but also produce the low-lying excitons. During the migration of hole, there will be spin conversion between the S and T states of electron-hole pair according to the hyperfine interactions in the DD pair mentioned above. On contact of the electron and the hole, spin of the pair may not be the same as original one. Since the yield of the recombination product from the contact electron-hole pair depends on the ratio of spin character of the pair, the magnetic field effect on the spin conversion of the DD pair are observed both in the photocurrent and photoluminescence. Another pioneering work concerning with the DD pair mechanism is reported by Groff *et al* [107]. They observed the ultralow field effect on the delayed fluorescence from dye-sensitized anthracene crystal. This is because the enhancement of spin conversion in the field region less than the hyperfine coupling constant illustrated in Fig. 30. Not only in crystalline semiconductor of small molecules but also in disordered semiconductor of polymer, the magnetic field effects on photocurrent and photoluminescence based on the DD pair mechanism have been observed by different groups [108-115]. All the phenomena reported in those literatures can be basically understood by the coherent hyperfine interactions of the DD pair in the regime-A. It should be noteworthy that magnetoconductance effect due to the DD pair mechanism at the crossline-R region has been also detected in a photoconductive polymer films doped with electron acceptors (Fig. 31) [94,116]. Even though the spin relaxation in organic semiconductors is thought to be long, which comes from the light chemical elements involved and one of the advantages superior to inorganic semiconductors, there is a unique report of spin relaxation induced OMAR effect in a polymer system [117]. The spin relaxation of the DD pair increases the OMAR effect dramatically as shown in Fig. 31. Therefore, control of the incoherent spin dynamics of the DD pair is of importance from the viewpoint of application of molecular spintronics and it is desirable to elucidate the mechanism for the relaxation in organic semiconductor.

History of the research on the magnetic field effects on dark conductivity in organic semiconductor may go back to a paper of anthracene crystal published at 1967 [118] after the first report of photoconductivity [99]. Magnetic field effects on electroluminescence from crystals of anthracene [119] and tetracene [120] were also investigated. The observed magnetic field effects were originally interpreted in terms of triplet-charge interaction that is classified to the DT pair mechanism rather than the DD pair. Carrier injection from electrodes derives more complicated situations compared with the case of photoinjection. The DD pair generated by the electrode injection is nongeminate pair that has no spin correlation initially and large amount of triplet exciton can be generated by recombination of the DD pair. In addition to the DD pair, therefore, the chance to form the DT and TT pairs also increases. No MR effect was reported until the appearance of the papers of organic light-emitting diodes made from the small molecule $Alq_3$ [121,122]. Kalinowski *et al.* interpreted the observed MR and magnetoluminescence effects in terms of the DD pair mechanism. On the other hand, Wohlgenannt *et al.* intensively investigated the OMAR effect using various semiconductors of polymers as well as small molecules [123-125]. It was found that the line width of OMAR nearly corresponding to $B_{1/2}$ is of the order of 5 mT. This value is reasonable as a typical $B_{1/2}$ due to the hyperfine interaction in organic molecules, but several models of triplet exciton-polaron interaction [126-131], electron-hole pair [132-136] and bipolaron [137] models are proposed to interpret the OMAR. The mechanism for OMAR is still controversial. The triplet exciton-polaron interaction model is a kind of DT pair mechanism. A recent review article [138] summarizes many experimental examples of the exciton-polaron model very well. Electron-hole pair and bipolaron models are classified to the DD pair mechanism. In the electron-hole pair model, the charge recombination depends on the spin state of the DD pair so that the carrier density is sensitive to the magnetic field, while OMAR on the bipolaron model stands on ideas of the mobility difference between polaron and bipolaron, and the bipolaron formation sensitive to magnetic field. Since the charge recombination is a ubiquitous carrier dynamics in organic semiconductor the electron-hole pair model is widely applicable for OMAR, but it is impossible to fit in unipolar devices [139,140]. As illustrated in Fig. 32, in bipolaron model, the DD pair is a polaron pair of electron-electron or hole-hole. Because triplet bipolaron has a very higher energy than the singlet bipolaron due to strong on-site exchange interaction, the formation of triplet bipolaron from the triplet polaron pair is forbidden energetically. The bipolaron model succeeds to reproduce many experimental OMAR with positive and negative signs [136,141]. In OMAR for dark conductivity, although OMAR effect due to the DD pair mechanism on the crossline-R was proposed theoretically [142], there is no clear experimental example yet. On the other hand, the ultralow field effect has been reported very recently [91,143]. Progress of fundamental researches using various sorts of materials and devices introduced in this section has been revealing the mechanism of OMAR that is the most important issue we must solve for development of technology of molecular spintronics.

## 5. Summary and future prospects

The authors introduced the current status of molecular spintronics. Spin injection, spin-dependent transport and spin-dependent binding effects, yielding magnetoresistance, were described in detail by introducing important achievements in the history of molecular spintronics. Several theories, which are necessary to carry out studies of molecular spintronics, were also described. As future prospects in this attractive new research field, because graphene is currently a model material, spin transistors and/or reconfigurable logic circuits will be fabricated in near future, which is an important milestone. Realization of long spin coherence with reproducibility and reliability is also strongly demanded. In addition, new experimental approaches for generating a spin current in molecules without using electric spin injection techniques should be proposed in order to establish novel methods for spin transport in molecules. As for the spin-dependent binding effects, the first observation of the magnetoresponse of conductivity and emission of organic device without magnetic compounds and electrodes might be a sort of discovery. However the many researches concerning about this intrinsic property in molecular semiconductors so far have been elucidating involvement of the complex paramagnetic pair dynamics with the spin-dependent binding process. The spin angular momentum is basically a conservation number during electron motion. Parallel to the history of fast-growing spintronics using metal and semiconductor, the organic is likely to lead to another new revolution. The intrinsic magnetoresponse in molecular semiconductors has an excellent advantage for molecular spintronics, because it happens even at room temperature and low fields and there is no need to use magnetic materials in device. Also these mild conditions permit a quick application to practical organic electronics. For enhancement of the performance and multi-functionalization of devices such as organic light emitting diode, organic photovoltaic cell and so on, no big change of the devise structure and materials is needed for the application of the magnetic response. However the mechanism of this intrinsic effect is not obvious yet. This is a main reason why we do not obtain an effective engineering technology to magnify the efficiency of the magnetic response. The clarification of the mechanism of the magnetoresponse is the most urgent and crucial subject. The ultralow field and spin relaxation effects would be a key issue to solve the fundamental problems. It is absolutely essential for physicists, chemists and engineers to meet in this interdisciplinary exciting research field.

# Appendix

As shown in eq. (11), the Dirac equation with an electromagnetic field can be written as

$$i\frac{\partial}{\partial t}\psi' = [e^{iS}(H - i\frac{\partial}{\partial t})e^{-iS}]\psi' = H'\psi'. \tag{11}$$

Here, the Unitary transformation can be written as,

$$\exp(+iS)H\exp(-iS) = H + i[S,H] + \frac{i^2}{2!}[S,[S,H]] + \cdots, \tag{A.1}$$

where the Baker-Hausdorff formula is used. Hence,

$$H' = H + i[S,H] - \frac{1}{2}[S,[S,H]] - \frac{i}{6}[S,[S,[S,H]]]$$
$$+ \frac{1}{24}[S,[S,[S,[S,\beta m]]]] - \dot{S} - \frac{i}{2}\left[S,\dot{S}\right] + \frac{1}{6}\left[S,\left[S,\dot{S}\right]\right]. \tag{A.2}$$

The above Hamiltonian is assumed to correspond to,

$$H' = \beta m + \varepsilon + \theta + i[S,\beta]m,$$

and the odd terms in H' should be eliminated. For this purpose, we set S as,

$$S = -\frac{i\beta\theta}{2m}. \tag{A.3}$$

Then, we carry out the following calculation, [S, H], [S, [S, H]], [S, [S, [S, H]]] and so on. Because $\beta$ and $\theta$ are anti-commutable and $\beta^2 = 1$, then

$$i[S,H] = i\left[-\frac{i\beta\theta}{2m}, \beta m + \theta + \varepsilon\right]$$
$$= -\theta + \frac{\beta}{m}\theta^2 + \frac{\beta}{2m}[\theta,\varepsilon],$$

$$\frac{i^2}{2!}[S,[S,H]] = \frac{i^2}{2}\left[S, i\theta - i\frac{\beta}{m}\theta^2 - i\frac{\beta}{2m}[\theta,\varepsilon]\right]$$
$$= -\frac{\beta}{2m}\theta^2 - \frac{\beta}{2m^2}\theta^3 - \frac{1}{8m^2}[\theta,[\theta,\varepsilon]],$$

$$\frac{i^3}{3!}[S,[S,[S,H]]] = \frac{1}{6m^2}\theta^3 - \frac{1}{6m^3}\beta\theta^4,$$

$$\frac{i^4}{4!}[S,[S,[S,[S,H]]]] = \frac{\beta\theta^4}{24m^3},$$

$$-\dot{S} = \frac{i\beta\dot{\theta}}{2m},$$

$$-\frac{i}{2}\left[S,\dot{S}\right] = -\frac{i}{8m}\left[\theta,\dot{\theta}\right].$$

Hence,

$$H' = \beta(m + \frac{\theta^2}{2m} - \frac{\theta^4}{8m^3}) + \varepsilon - \frac{1}{8m^2}[\theta,[\theta,\varepsilon]] - \frac{i}{8m^2}\left[\theta,\dot{\theta}\right]$$
$$+ \frac{\beta}{2m}[\theta,\varepsilon] - \frac{\theta^3}{3m^2} + \frac{i\beta\dot{\theta}}{2m} \equiv \beta m + \varepsilon' + \theta', \qquad (A.4)$$

where it is noteworthy that the odd term is included only in the term of $O(1/m)$. In order to rewrite the above Hamiltonian as,

$$H'' = \beta m + \varepsilon' + \theta'', \qquad (\theta'' \sim O(\frac{1}{m^2}))$$

the Unitary transformation is once again implemented by setting $S'$ as,

$$S' = \frac{-i\beta}{2m}\theta' = \frac{-i\beta}{2m}(\frac{\beta}{2m}[\theta,\varepsilon] - \frac{\theta^3}{3m^2} + \frac{i\beta\dot{\theta}}{2m}). \qquad (A.5)$$

Then the Hamiltonian can be rewritten as,

$$H'' = \beta m + \varepsilon' + \frac{\beta}{2m}[\theta',\varepsilon'] + \frac{i\beta\dot{\theta'}}{2m}$$
$$\equiv \beta m + \varepsilon' + \theta''. \qquad (A.6)$$

Furthermore, we set $S''$ as,

$$S' = \frac{-i\beta}{2m}\theta'', \qquad (A.7)$$

and thus,

$$H'' = \beta(m + \frac{\theta^2}{2m} - \frac{\theta^4}{8m^3}) + \varepsilon - \frac{1}{8m^2}[\theta,[\theta,\varepsilon]] - \frac{i}{8m^2}\left[\theta,\dot{\theta}\right], \qquad (A.8)$$

is the transformed Hamiltonian, where the odd term becomes small enough ($\sim O(1/m^2)$). The third and the fourth terms in eq. (A.8) can be written as,

$$-\frac{1}{8m^2}\left[\theta,[\theta,\varepsilon]+i\dot{\theta}\right],$$

and we carry out the calculation of the above term explicitly, as follows;

$$\frac{\theta^2}{2m}=\frac{\{\alpha\cdot(p-eA)\}^2}{2m}=\frac{1}{2m}\{(p-eA)^2+i\alpha\cdot(p-eA)\times(p-eA)\}$$

$$=\frac{1}{2m}\{(p-eA)^2-e(\sigma\cdot B)\},$$

$$\frac{1}{8m^2}([\theta,\varepsilon]+i\dot{\theta})=\frac{1}{8m^2}([\alpha\cdot(p-eA),e\phi]+i\dot{\theta})$$

$$=\frac{1}{8m^2}(-ie\alpha\nabla\phi-ie\alpha\dot{A})$$

$$=\frac{ie}{8m^2}\alpha\cdot E$$

$$\left[\theta,\frac{i}{8m^2}\alpha\cdot E\right]=\frac{ie}{8m^2}[\alpha\cdot p,\alpha\cdot E]$$

$$=\frac{ie}{8m^2}\sum_{i,j}\alpha_i\alpha_j(-i\frac{\partial E^j}{\partial x^i})+\frac{e}{4m^2}\sigma(E\times p)$$

$$=(\frac{e}{8m^2}\sigma\cdot\text{rot}(E)+\frac{ie}{8m^2}\text{div}(E))+\frac{e}{4m^2}\sigma(E\times p).$$

In the above calculation, remember that the time derivative of momentum in the Schroedinger representation is equal to zero. From above calculations, we can obtain eq. (12),

$$H'=\beta\{m+\frac{(p-eA)^2}{2m}-\frac{p^4}{8m^3}\}+e\phi-e\frac{1}{2m}\beta(\sigma\cdot B)$$

$$-\frac{ie}{8m^2}\sigma\cdot\text{rot}(E)-\frac{e}{4m^2}\sigma\cdot(E\times p)-\frac{e}{8m^2}\text{div}(E). \qquad (12)$$

**Figures and figure captions**

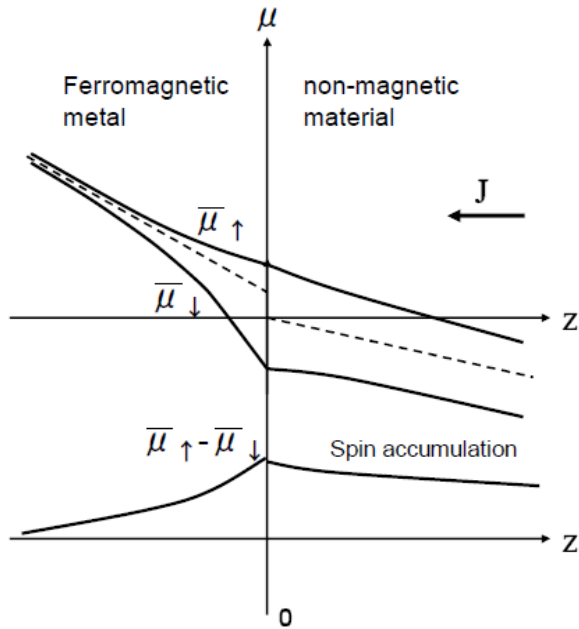

Fig. 1 A Schematic diagram of the electrochemical potential for up- and down-spin in a ferromagnet/nonmagnet heterojunction.

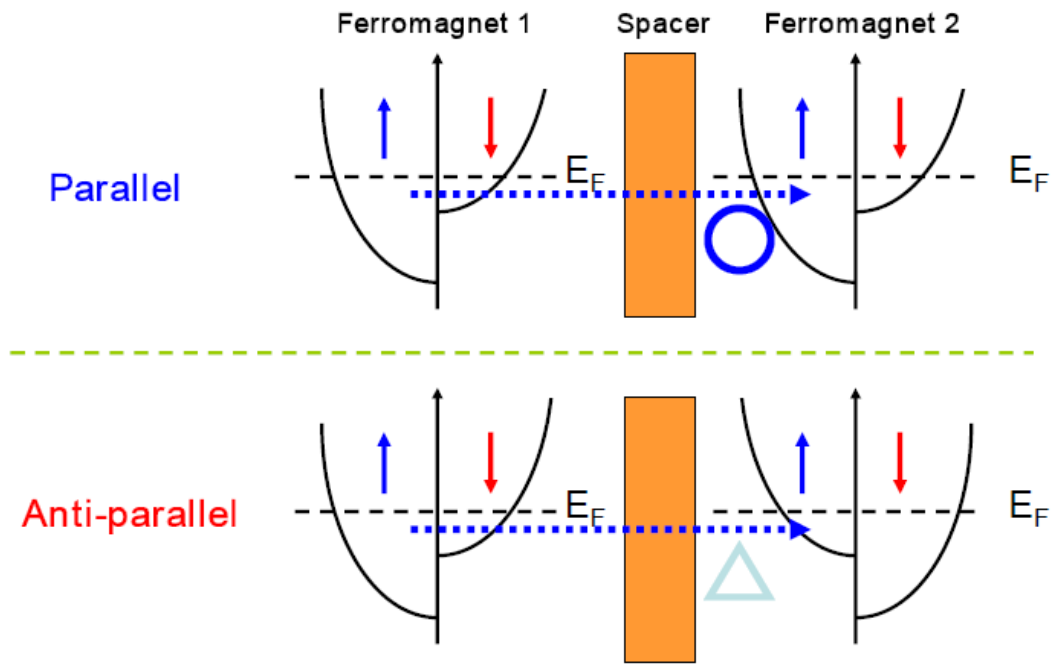

Fig. 2 A schematic diagram of spin-dependent transport. Here, the spacer plays a role of tunneling barrier. The DOS for the up spin in ferromagnet 2 changes in parallel and anti-parallel magnetization configurations.

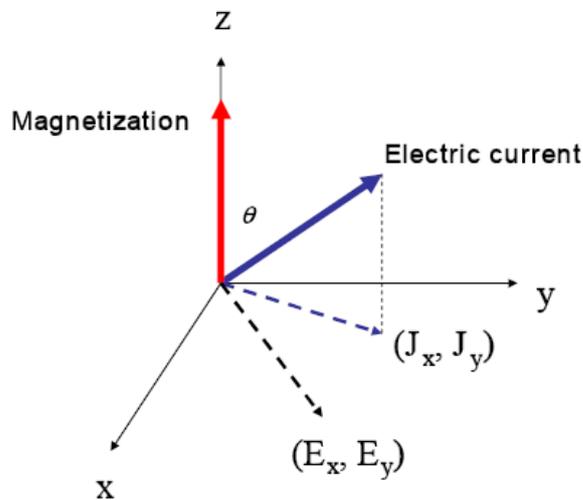

Fig. 3 A direction relationship between the magnetization and the electric current in the case of anisotropic MR effect.

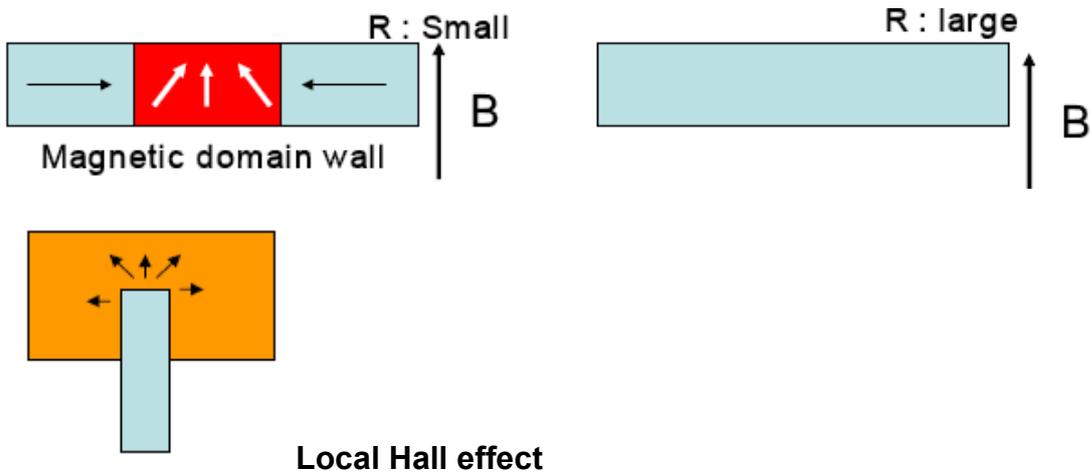

**Local Hall effect**

Fig. 4 Various origins of spurious signals. The top panel shows anisotropic MR effect, which is attributed to the generation of a magnetic domain wall in a ferromagnetic electrode during magnetization reversal. The external magnetic field (B) is applied perpendicular to the ferromagnetic electrodes. The bottom panel shows a schematic of local Hall effect, which occurs when the spin transport channel is not completely covered by ferromagnet.

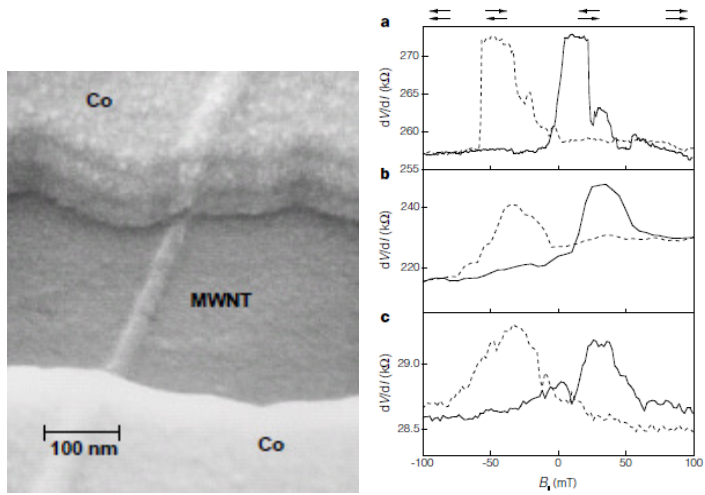

Fig. 5 The spin valve geometry (left) and resistance hysteresis observed at 4.2 K (right). From Tsukagoshi [19].

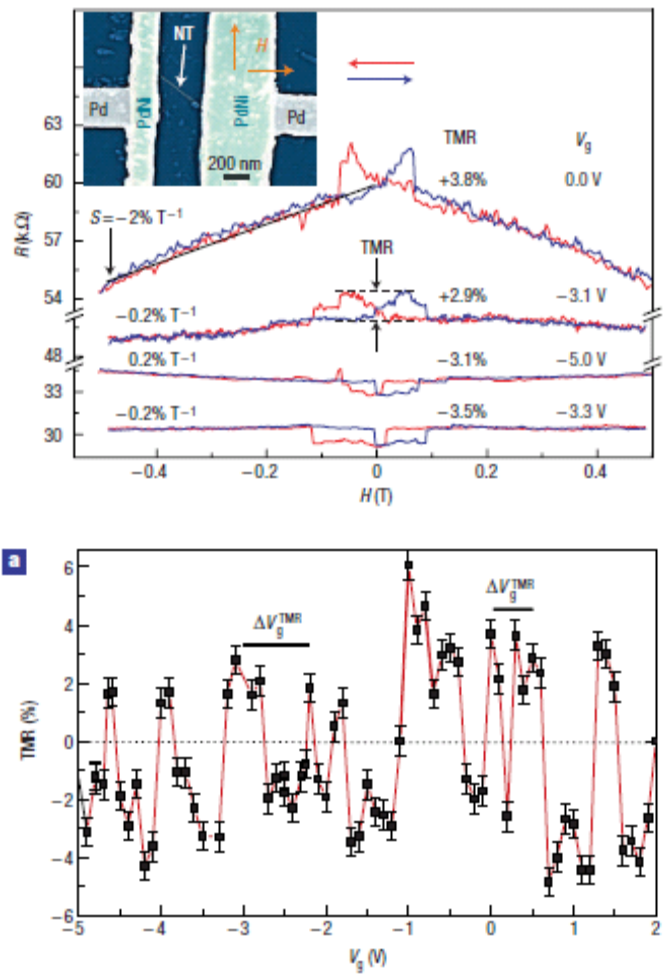

Fig. 6 (The upper) The observed MR effects in the PdNi-contacted nanotubes. The sign and the values of the MR ratio oscillated as the gate voltage was changed. (The lower) The oscillation of the

MR ratio. The black closed squares are experimental data and the red solid line shows the theoretical fitting. From Sahoo [25].

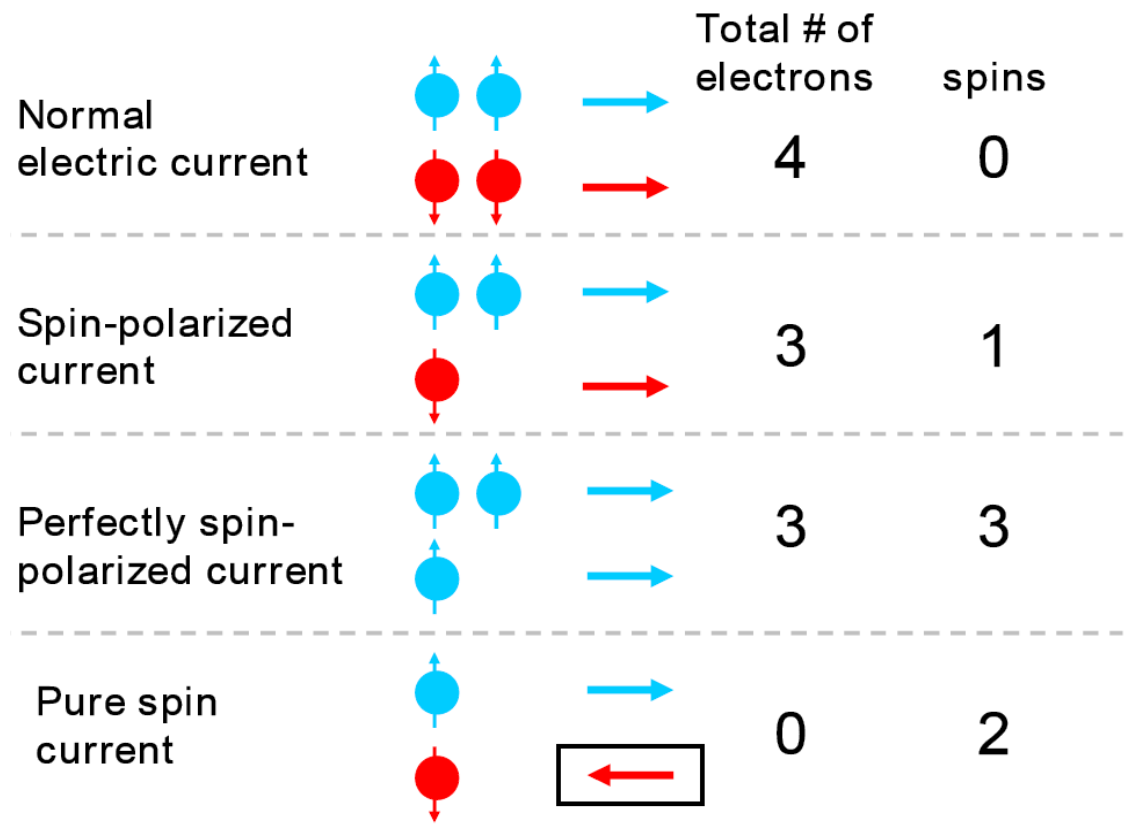

Fig. 7　Comparison of a normal electric current, a spin polarized current and a pure spin current.

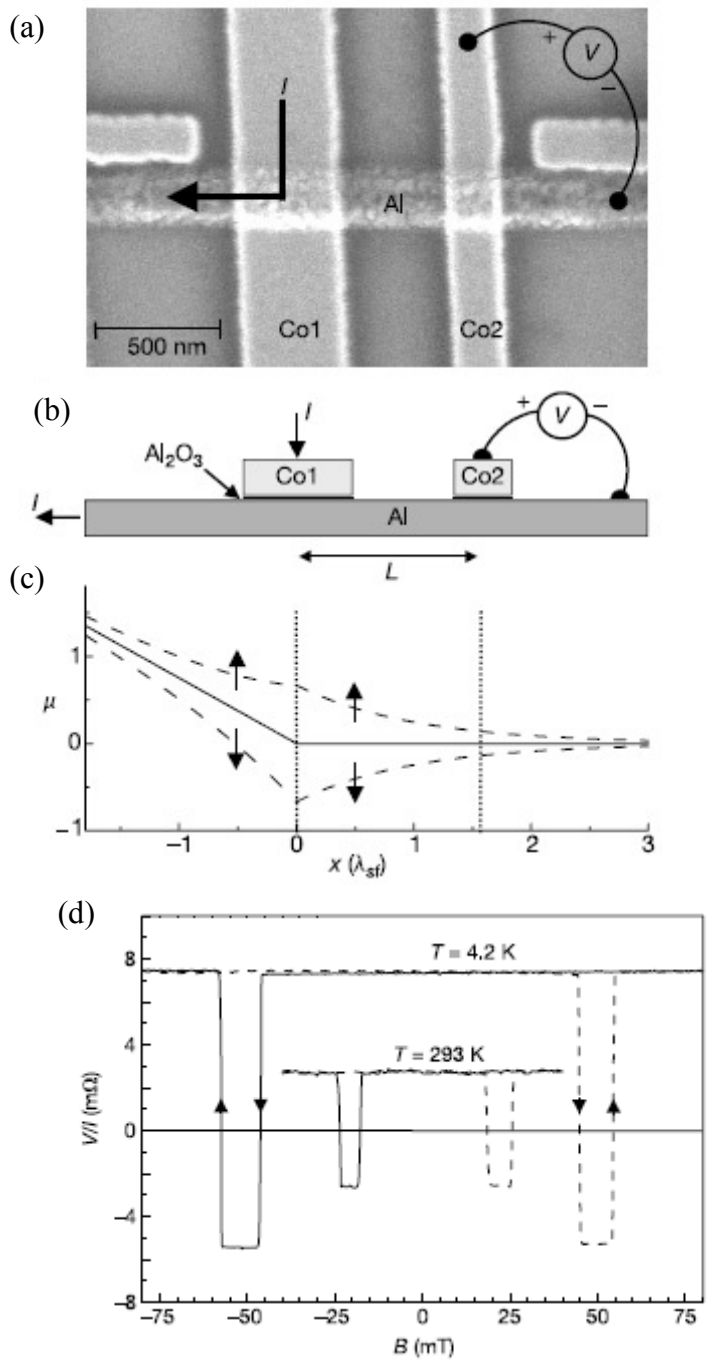

Fig. 8 (a) An scanning electron microscopic image of a Co/AlO/Al spin valve. (b) A cross-sectional schematic of the metallic spin valve. (c) Position dependence of electrochemical potential for up- and down-spin. (d) Observed spin signals at 4.2 and 293 K. From Jedema [28].

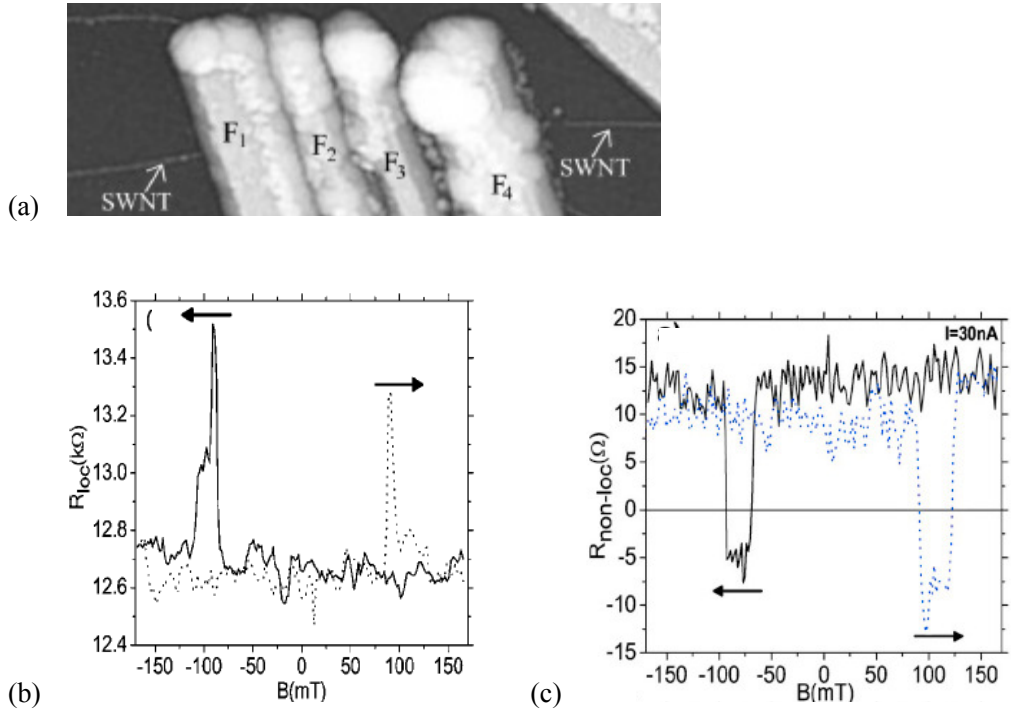

Fig. 9 (a) An scanning electron microscopic image of a SWNT spin valve with four ferromagnetic electrodes. (b) A spin signal in the local geometry at 4.2 K (Injection current was 10 nA). (c) A spin signal in the non-local geometry at 4.2 K (Injection current was 30 nA). From Tombros [27].

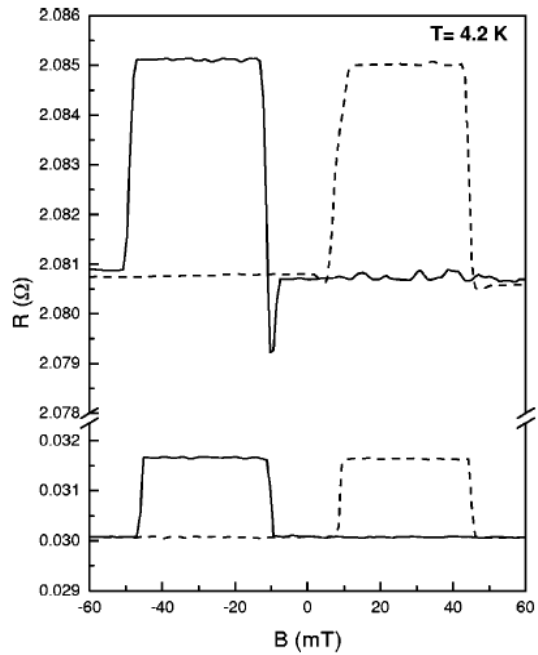

Fig. 10 Spin signals in the conventional "local" method (top curve) and "non-local" method (bottom curve) observed in Py spin valves at 4.2 K. The ratio of the signal intensities is equal to 2. From Jedema [30].

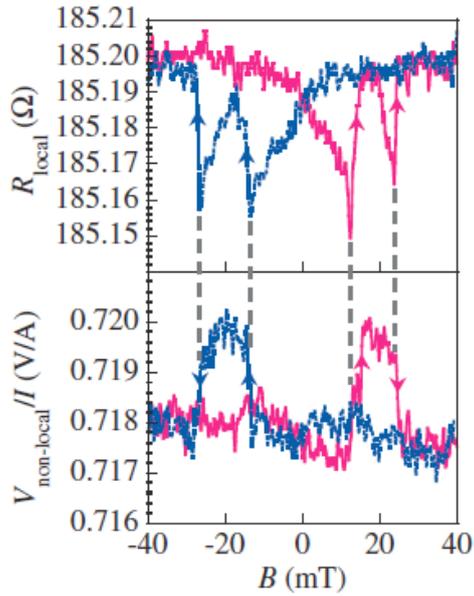

Fig. 11 (Top panel) MR signal due to AMR effect observed in a MLG spin valve at RT. (Bottom panel) Spin injection signal in a MLG spin valve at RT. From Ohishi [37].

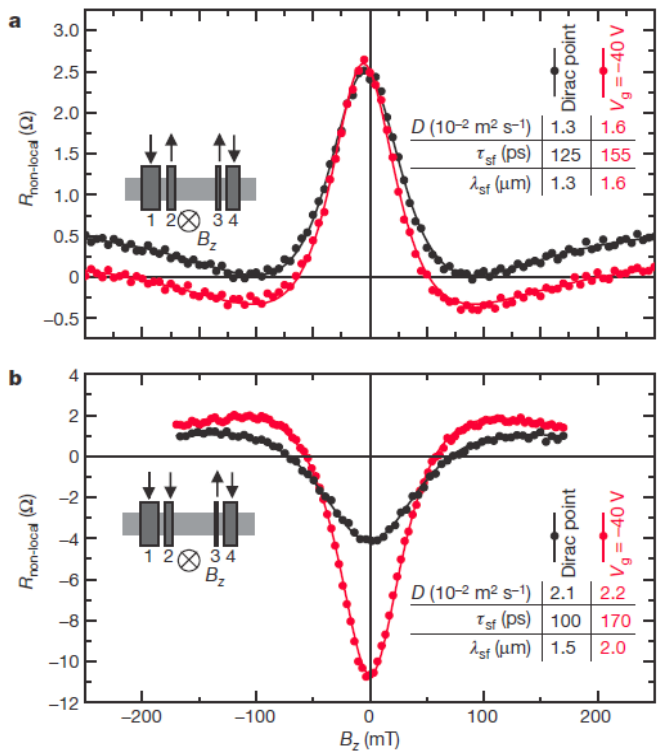

Fig. 12 Hanle-type spin precession observed in SLG. (a) The precession in the parallel alignment. (b) The precession in the anti-parallel alignment. The red line is for Vg=-40V, and the black line is for the Dirac point. From Tombros [38].

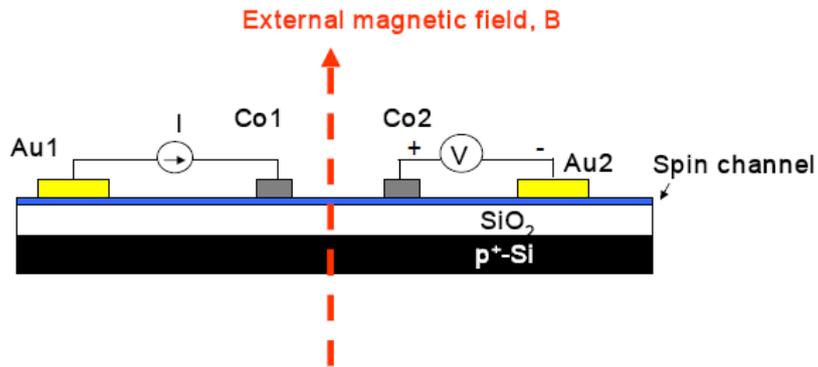

Fig. 13 A schematic of observing the Hanle-type spin precession. The external magnetic field is applied perpendicular to the spin channel.

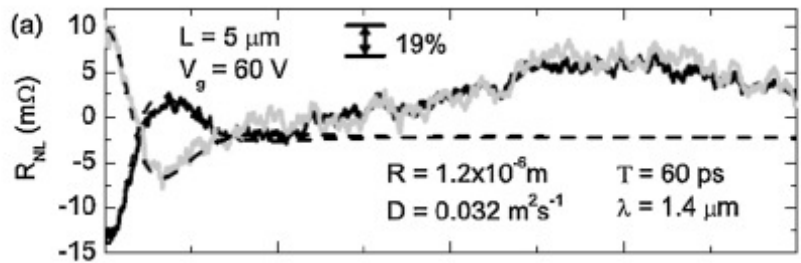

Fig. 14 An example of anisotropic spin relaxation in SLG. From Tombros [42].

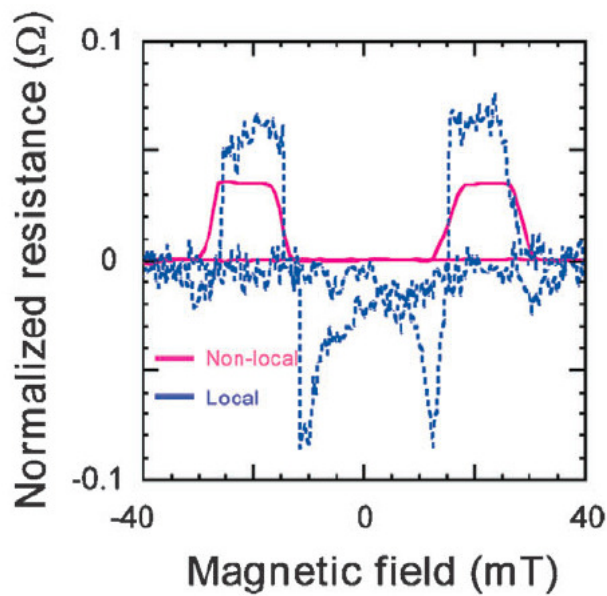

Fig. 15 Comparison of the spin signal intensities observed in MLG at RT in the local and the non-local geometries. The ratio of the intensities is almost equal to two. From Shiraishi [40].

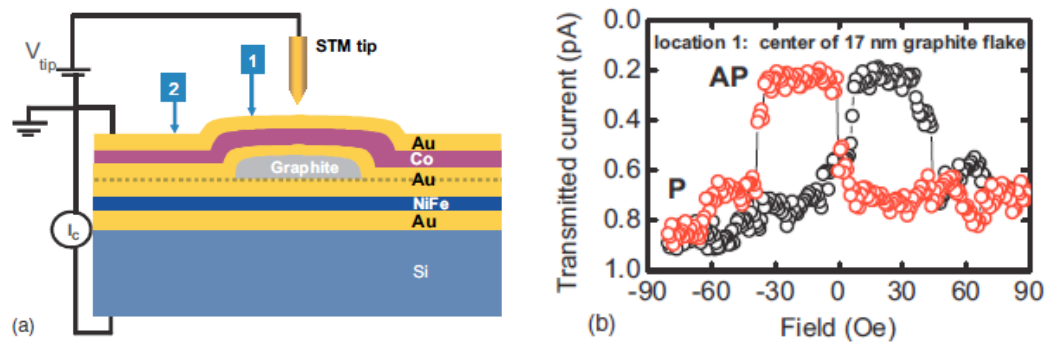

Fig. 16 (a) A schematic of the BEEM technique. (b) The observed local spin-valve signals at 150 K. From Benerjee [53].

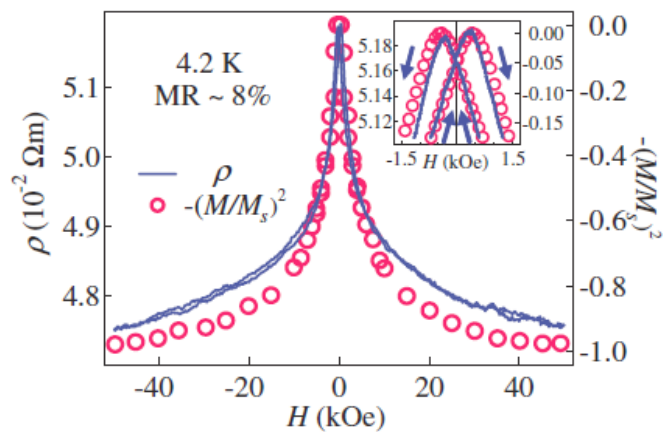

Fig. 17 The observed MR curve (the solid line) and the corresponding magnetization of the Co (open circles) in a $C_{60}$-Co nano-composite spin device. From Miwa [55].

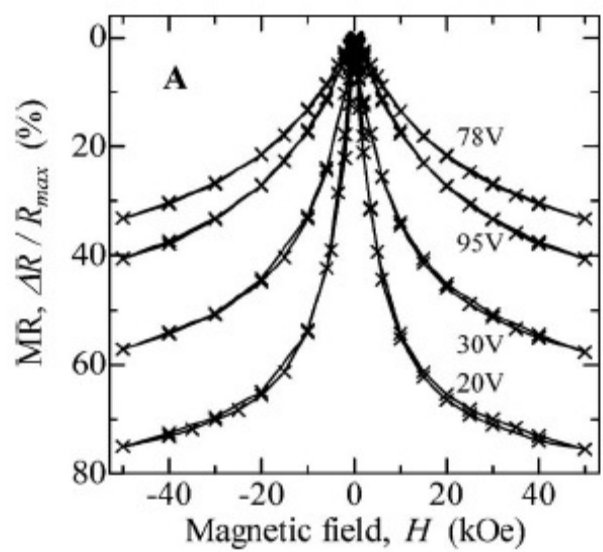

Fig. 18 An example of the large MR ratio in C60-Co at 4.2 K. From Sakai [57].

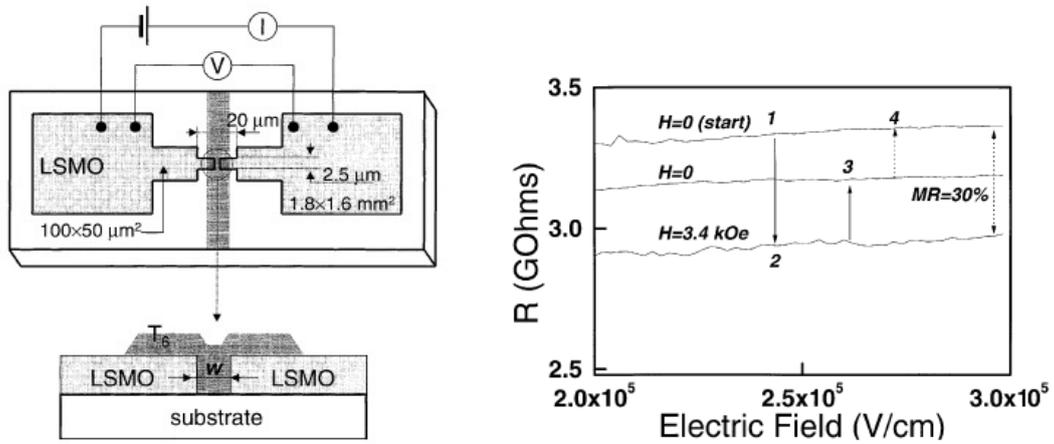

Fig. 19 (Left) Schematics of an organic spin valve using T6 and LSMO. (Right) The observed resistance changes as a function of the external magnetic field. From Dediu [60]

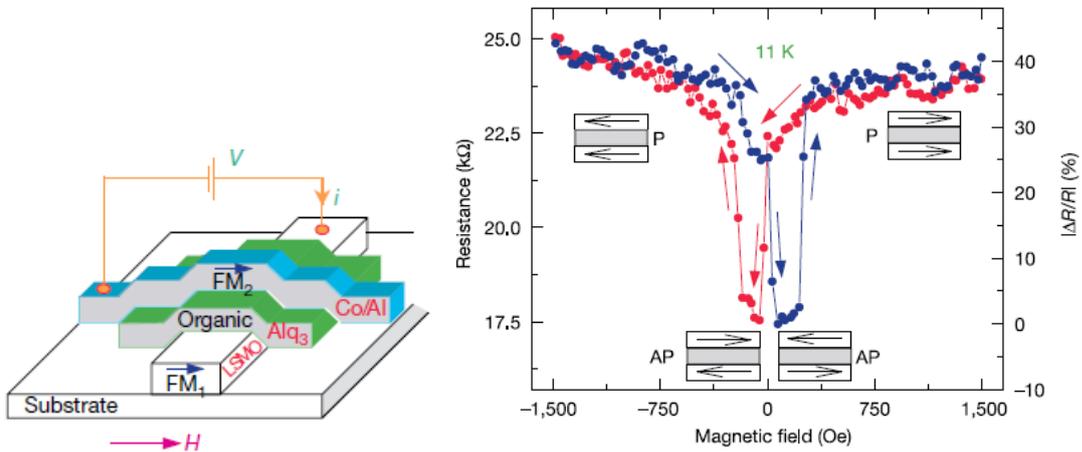

Fig. 20 (Left) A schematic of a vertical spin valve using Alq3. (Right) The observed MR effect in the organic spin valve at 11 K. From Xiong [62].

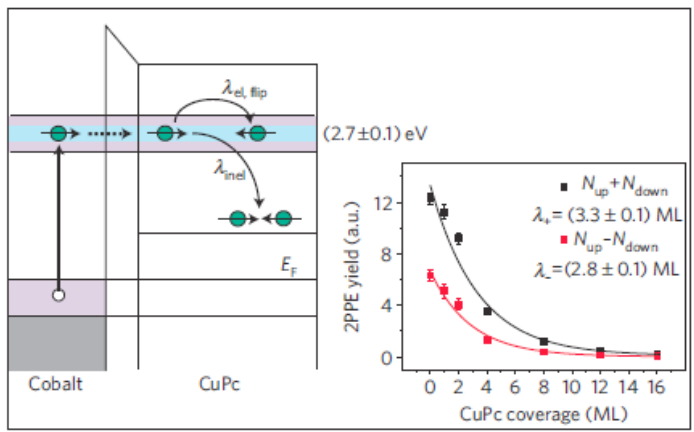

Fig. 21 A result on the 2PPE experiment in the purely spin injected case. From Cinchetti [67].

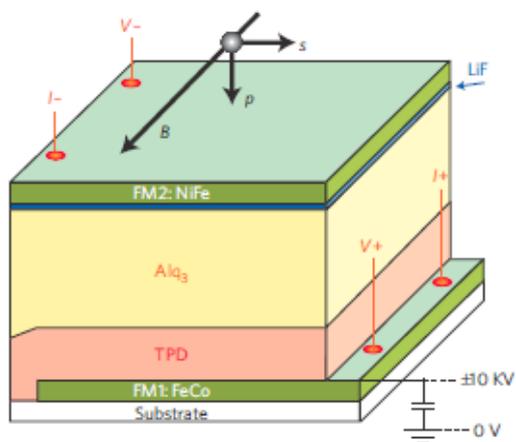
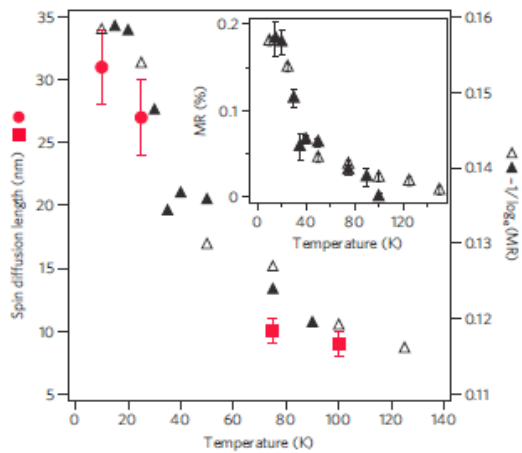

Fig. 22 (Left) The spin valve structure. (Right) The estimated spin diffusion length as a function of temperature. From Drew [68].

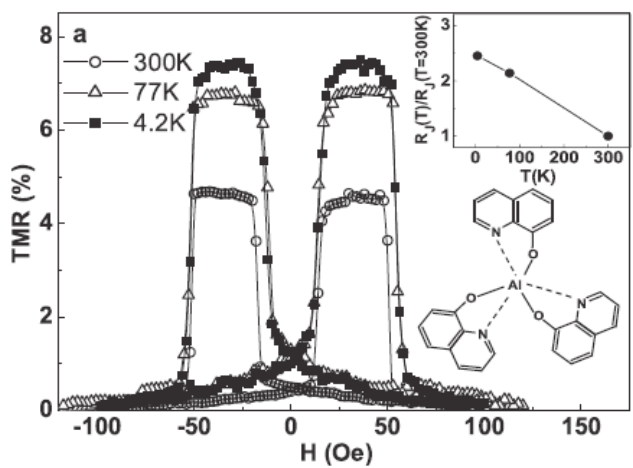

Fig. 23 An example of spin-dependent tunneling transport via molecules (here Alq3) in a Py/Alq3/Co spin valve. From Santos [69].

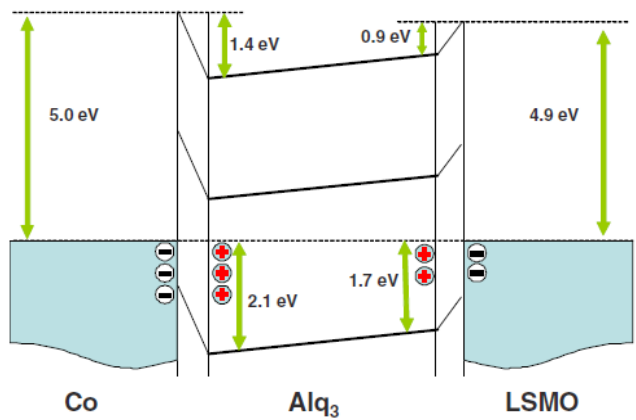

Fig. 24 A band diagram in a Co/Alq3/LSMO spin valve, which was obtained by using photoemission spectroscopy. From Zhan [77].

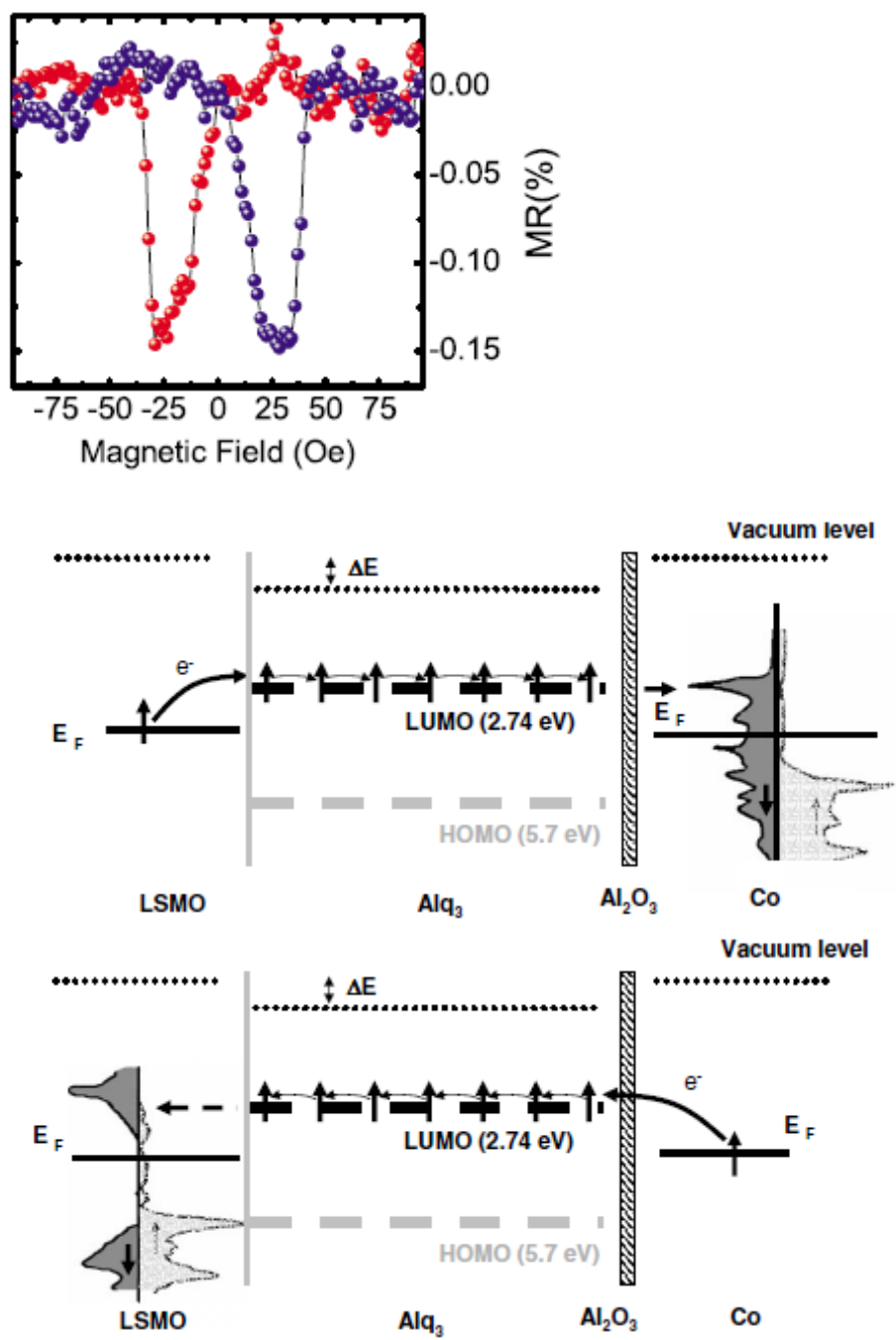

Fig. 25 (Top) The observed MR effect in a Co/AlO/Alq3/LSMO spin valve at RT. (Bottom) A model of spin transport in the Alq3. The injected spins transfer in the LUMO level of the Alq3 both in the case that the spins are injected from Co and from LSMO. This is thought to be a reason of the negative MR effect. From Dediu [78].

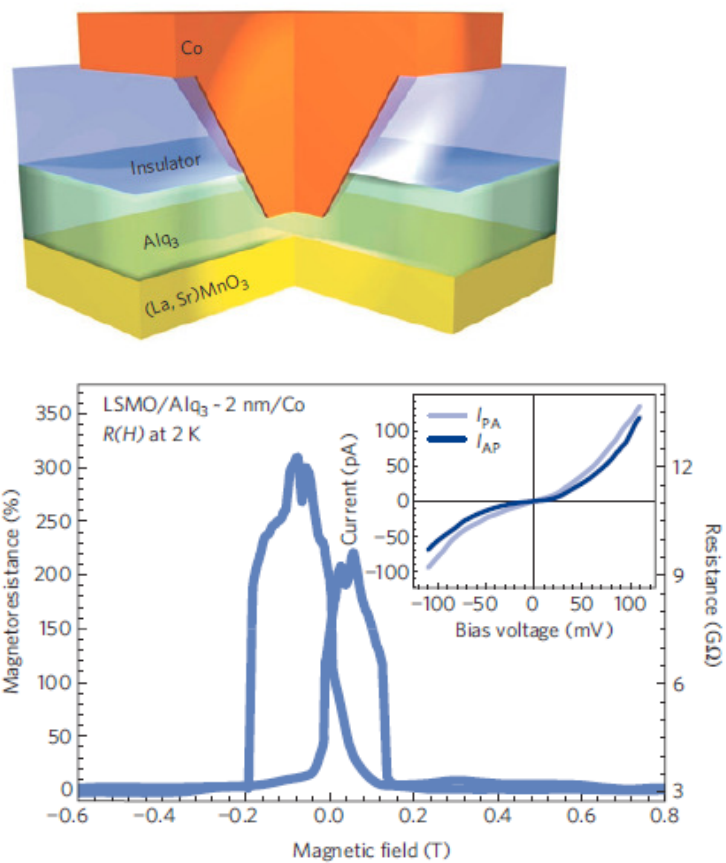

Fig. 26 (Left) A schematic of a Alq3 spin valve fabricated by a nano-indentation method. (Right) The observed MR effect at 2 K. From Barraud [82].

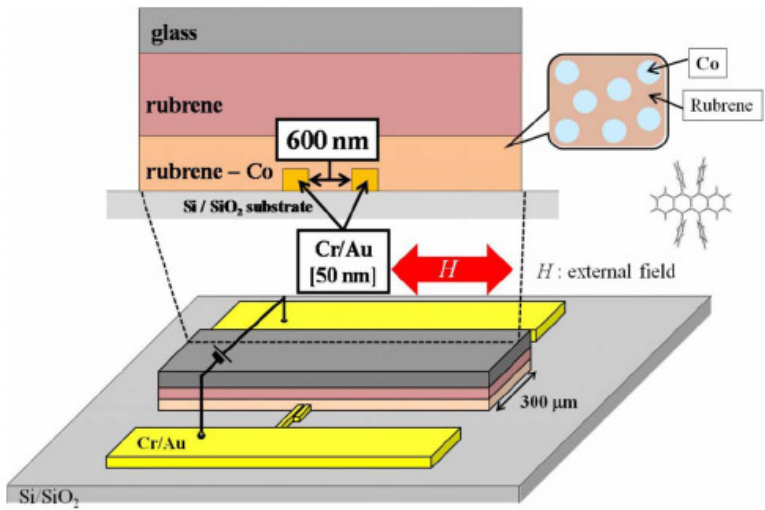

(a)

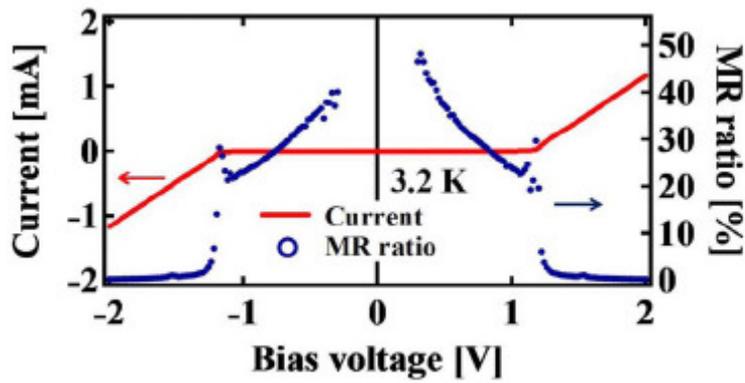

(b)

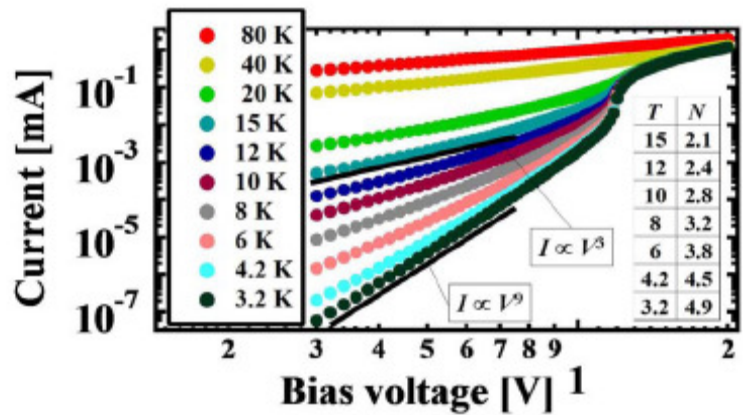

(c)

Fig. 27 (a) A device structure of a rubrene-cobalt nano-composite spin device. (b) Correspondence between the *I-V* curve and the MR ratio at 3.2 K. (c) The *I-V* curves within the Coulomb gap. $V^{2N-1}$ dependence was observed. From Hatanaka [84].

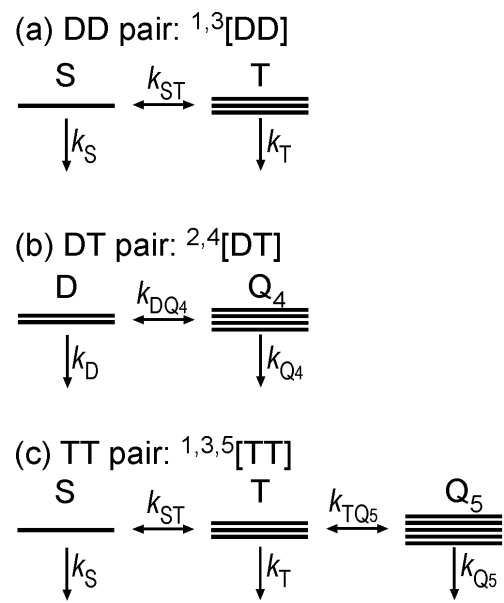

Fig. 28 Electron spin states and dynamics of paired particles with unpaired electrons.

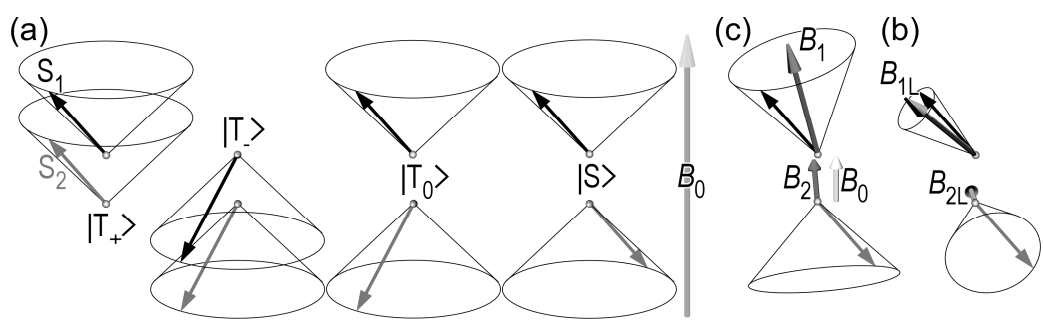

Fig. 29 Vector representation of the DD pair under (a) strong, (b) zero and (c) weak external magnetic fields.

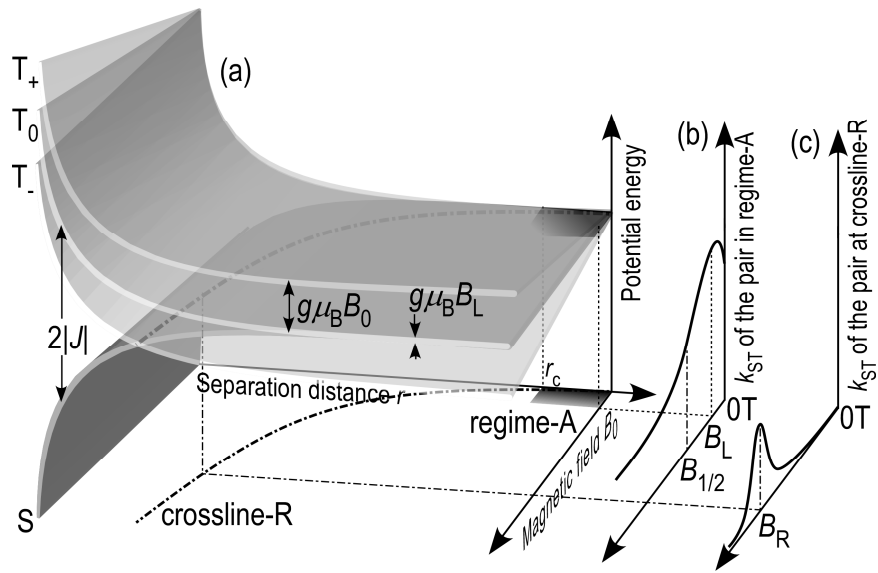

Fig. 30 Separation distance and magnetic field dependences of the potential energy for spin sublevels of the DD pair. The electrostatic interaction between the D-particles is excluded in the potential energy for simplicity.

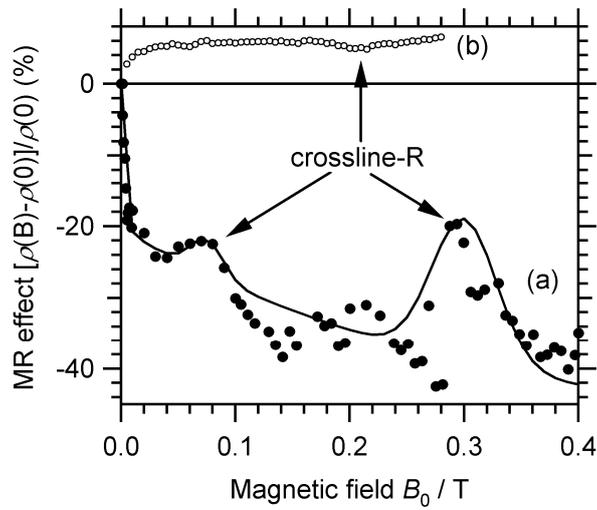

Fig. 31 MR effects for the photoconductivity of poly(N-vinylcarbazole) films doped with lumichrome (a) and 1,3,4,5-tetracyanobenzene (b) observed by selective excitation under ambient conditions. The negative and positive MRs respectively arises from magnetic field effects on the singlet and triplet electron-hole pairs. The $^{1,3}$excitons generate geminate $^{1,3}$DD pair. Partially from Ikoma [115].

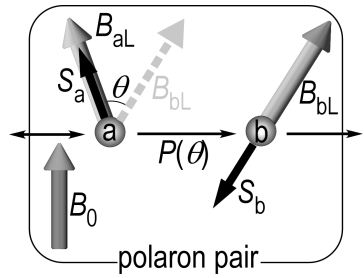

Fig. 32 Model of bipolaron formed from the pair consisting of two polarons with 1/2-electron spins ($S_a$ and $S_b$) at site a and b. The polarons respectively have different effective magnetic fields $B_{a,b}$, which come from individual hyperfine fields. Polaron transfer from site a to site b stands for binding to bipolaron. The bipolaron formation probability $P(\theta)$ depends on the singlet character of the polaron pair. The singlet character in the pair obeys $1/4 - |S_a||S_b|\cos\theta/\hbar^2$, in which $\theta$ is the angle between $B_a$ and $B_b$ and approaches zero with the increase of the $B_0$ field